\documentclass[a4paper]{article}
\usepackage{graphicx}
\usepackage{dcolumn}
\usepackage{bm}
\usepackage{longtable}
\usepackage{comment}
\usepackage[normalem]{ulem}
\usepackage{amsfonts}
\usepackage{epstopdf}
\newcommand{\bea}{\begin{eqnarray}}
\newcommand{\eea}{\end{eqnarray}}
\newcommand{\ba}{\begin{eqnarray}}
\newcommand{\ea}{\end{eqnarray}}

\newcommand{\be}{\begin{equation}}
\newcommand{\ee}{\end{equation}}

\def\Be'{\beta_\mu^{'}}

\def\<{\bigl\langle}
\def\>{\bigr\rangle}

\providecommand{\be}{\begin{equation}}
  \providecommand{\ee}{\end{equation}}
\providecommand{\bea}{\begin{eqnarray}}
  \providecommand{\eea}{\end{eqnarray}}
\providecommand{\beas}{\begin{eqnarray*}}
  \providecommand{\eeas}{\end{eqnarray*}}

\providecommand{\beni}{\begin{equation*}}
  \providecommand{\eeni}{\end{equation*}}

\providecommand{\bw}{\begin{widetext}}
  \providecommand{\ew}{\end{widetext}}

\providecommand{\sign}{\mathop{\mathrm{sign}}}
\usepackage{amssymb,amsfonts,amsmath,graphics,graphicx}

\title{Hierarchical neural networks perform both serial and parallel processing}

\author{Elena Agliari  \footnote{
    Dipartimento di Fisica, Sapienza Universit\`a di Roma, P.le A. Moro 2, 00185, Roma}, Adriano Barra  \footnote{
    Dipartimento di Fisica, Sapienza Universit\`a di Roma, P.le A. Moro 2, 00185, Roma}, Andrea Galluzzi
    \footnote{Dipartimento di Matematica, Sapienza Universit\`a di Roma, P.le A. Moro 2, 00185, Roma},\\ Francesco Guerra  \footnote{Dipartimento di Fisica, Sapienza Universit\`a di Roma, P.le A. Moro 2, 00185, Roma}, Daniele Tantari  \footnote{Dipartimento di Matematica, Sapienza Universit\`{a} di Roma, P.le A. Moro 2, 00185, Roma}, Flavia Tavani  \footnote{Dipartimento SBAI (Ingegneria), Sapienza Universit\`{a} di Roma, Via A. Scarpa 14, 00185, Roma}}

\begin{document}
\maketitle

\begin{abstract}
In this work we study a Hebbian neural network, where neurons are arranged according to a hierarchical architecture such that their couplings scale with their reciprocal distance.
As a full statistical mechanics solution is not yet available, after a streamlined introduction to the state of the art via that route, the problem is consistently approached through signal-to-noise technique and extensive numerical simulations. Focusing on the low-storage regime, where the amount of stored patterns grows at most logarithmical with the system size, we prove that these non-mean-field Hopfield-like networks display a richer phase diagram than their classical counterparts. In particular, these networks are able to perform serial processing (i.e. retrieve one pattern at a time through a complete rearrangement of the whole ensemble of neurons) as well as parallel processing (i.e. retrieve several patterns simultaneously, delegating  the management of different patterns to diverse communities that build network). The tune between the two regimes is given by the rate of the coupling decay and by the level of noise affecting the system.
\newline
The price to pay for those remarkable capabilities lies in a network's capacity  smaller than the mean field counterpart, thus yielding a new budget principle: the wider the multitasking capabilities, the lower the network load and viceversa. This may have important implications in our understanding of biological complexity.
\end{abstract}


\section{Introduction}\label{sec:I}

Statistical mechanics constitutes a powerful technique for the understanding of neural networks \cite{amit,peter}, however overcoming the mean-field approximation is extremely hard (even beyond neural networks). Basically, the mean-field approximation lies in assuming that each spin/neuron $S_i$  in a network dialogues with {\em all} the other spin/neurons with the same strength \footnote{Notice that this situation corresponds to a system embedded in a fully-connected (i.e. complete graph) topology. However, situations where we introduce some degree of dilution (e.g. Erd\"{o}s-R\'{e}nyi graph), yet preserving the homogeneity of the structure and an extensive coordination number, can be looked and treated as mean field models.}. For instance, if we consider a ferromagnetic model, once introduced $N$ spins $S_i = \pm1$, $i \in (1,...,N)$, we have the two extreme scenarios of a nearest-neighbor model like the Ising lattice, whose Hamiltonian can be written as
\be
H_{\textrm{Ising}} = -\sum_{\langle i,j \rangle }J S_i S_{j},
\ee
where, crucially, the sum runs over all the couples $\langle i,j \rangle$ of {\em adjacent} sites, and the mean-field Curie-Weiss model, whose Hamiltonian can be written as
\be\label{CW}
H_{\textrm{Curie-Weiss}} = -\sum_{i<j}^{N,N} J S_i S_j,
\ee
where the sum runs over {\em all} the $N(N-1)/2$ spin couples irrespective of any notion of distance; this is equivalent to think of spins interacting through nearest neighbor prescriptions but as they were embedded in an $N$-dimensional space. Clearly, solving the statistical mechanics of the latter model is much simpler with respect to the former. The main route toward finite-dimensional descriptions has been paved by physicists in the study of condensed matter\footnote{In that context the long-range interactions are unacceptable because the involved couplings are of electromagnetic nature, hence displaying power-law decay with the distance.}. Indeed, incredible efforts have been spent from the $70$s in working out the renormalization-group \cite{W1}, namely a technique which allows inferring the properties of three-dimensional ferromagnets starting from mean-field descriptions, but a straight solution of the Ising model in dimensions $3$ is still out of the current mathematical reach\footnote{It is worth mentioning that the Wilson-Kadanoff renormalization equations \cite{W2,W3,W4} turn out to be exact in models with power law interactions as those built on the hierarchical lattice that we are going to consider.}.

Actually, in the last decade some steps forward toward {\em more realistic} systems have been achieved merging statistical mechanics \cite{Gallavotti,ellis,MPV} and graph theory \cite{bollobas,barabasi,SW}. In particular, mathematical methodologies were developed to deal with spin systems embedded in random graphs, where the ideal, full homogeneity among spins is lost \cite{ton1,ton2}. Thus, networks of neurons arranged according to Erd\"{o}s-R\'enyi \cite{jstato}, small-world \cite{agliari2}, or scale-free \cite{tonno} topologies were addressed, yet finite-dimensional networks were still out of debate.

Focusing on neural networks, it should be noted that, beyond the difficulty of treating non-trivial topologies for neuron architecture, one has also to cope with the complexity of their coupling pattern, meant to encode the Hebbian learning rule. The emerging statistical mechanics is much  trickier than that for ferromagnets; indeed neural networks can behave either as ferromagnets or as spin-glasses, according to the parameter settings: their phase space is split into several disconnected pure states, each coding for a particular stored pattern, so to interpret the thermalization of the system within a particular energy valley as the spontaneous retrieval of the stored pattern associated to that valley. However in the high-storage limit, where the amount of patterns scales linearly with the number of neurons, neural networks approach pure spin-glasses (loosing retrieval capabilities at the blackout catastrophe \cite{amit}) and, as a simple Central Limit argument shows \cite{howglassyNN}, when the amount of patterns diverge faster that the amount of neurons they become purely spin glasses. For the sake of exhaustiveness we also stress that, even in the retrieval region, neural networks are {\em exactly} linear combinations of two-party spin glasses \cite{rsneuralanalog,bipCWSK}: due to the combination of such difficulties, neural networks on a finite dimensional topology have not been extensively investigated so far.

However, very recently, a non-mean-field model, where a topological distance among spins can be defined and couplings can be accordingly rescaled, turned out to be, to some extent, treatable also for complex systems such as spin-glasses \cite{castellana-pre,cecilia2}.  More precisely, spins are arranged according to a hierarchical architecture as shown in Fig.~$1$: each pair of nearest-neighbor spins form a ``dimer'' connected with the strongest coupling, then spins belonging to nearest ``dimers'' interact each other with a weaker coupling and so on recursively \cite{mukamel}. In particular, the Sherrington-Kirkpatrick model for spin-glasses defined on the hierarchical topology has been investigated in \cite{castellana-prl}: despite a full analytic formulation of its solution still lacks, renormalization techniques, \cite{castellana-pre,cecilia1}, rigorous bounds on its free-energies \cite{castellana-jsp} and extensive numerics \cite{leuzzi1,leuzzi2} can be achieved nowadays and they give extremely sharps hints on the thermodynamic  behavior of systems defined on these peculiar topologies.

Remarkably, as we are going to show, when implementing the Hebb prescription for learning on these hierarchical networks, an impressive phase diagram, much richer than the mean-field counterpart, emerges. More precisely, neurons turn out to be able to orchestrate both serial processing (namely sharp and extensive retrieval of a pattern of information), as well as parallel processing (namely retrieval of different patterns simultaneously).

The remaining of the paper is structured as follows: in the next subsections we provide a streamlined description of mean-field serial and parallel processors, and we introduce the hierarchical scenario. Then, we split in three sections our findings according to the methods  exploited for investigation: statistical mechanics, signal-to-noise technique and extensive numerical simulations. All these approaches consistently converge to the scenario outlined above. Seeking for clarity and completeness, each technique is first applied to a ferromagnetic hierarchical mode (which can be thought of as a trivial one-pattern neural network and acts as a test-case) and then for a low-storage hierarchical Hopfield model.

\subsection{Mean-field processing: Serial and parallel processors.}\label{sec:MF}

Probably the most famousmodel for neural networks is the Hopfield model presented in his seminal paper dated $1982$ \cite{AI1}, counting nowadays more than twenty-thousand citations (Scholar). This is a mean-field model, where neurons are schematically represented as dichotomic Ising spins (state $+1$ represents firing while state $-1$ stands for quiescence) interacting via a (symmetric rearrangement of) the Hebbian rule for learning as masterfully shown by the extensive statistical-mechanical analysis that Amit, Gutfreund and Sompolinsky performed on the model \cite{amit,AGS}.
\newline
More formally, once introduced $N$ neurons/spins $S_i$, $i \in (1,...,N)$, and $p$ quenched patterns $\bold{\xi}_{\mu}$, with $\mu \in (1,...,p)$, whose entries are drawn once for all from the uniform distribution
\be \label{eq:xi}
P(\xi_i^{\mu}) = \frac12 \delta(\xi_i^{\mu}-1) + \frac12 \delta(\xi_i^{\mu}+1),
\ee
the Hopfield model is then captured by the following Hamiltonian $H_{\textrm{Hopfield}}(S|\xi)$
\be\label{HMF}
H_{\textrm{Hopfield}}(S|\xi)= - \frac{1}{N}\sum_{i<j}^N \left( \sum_{\mu=1}^p \xi_i^{\mu}\xi_j^{\mu}\right)S_iS_j.
\ee
\begin{figure}[tb] \begin{center}
\includegraphics[width=.68\textwidth]{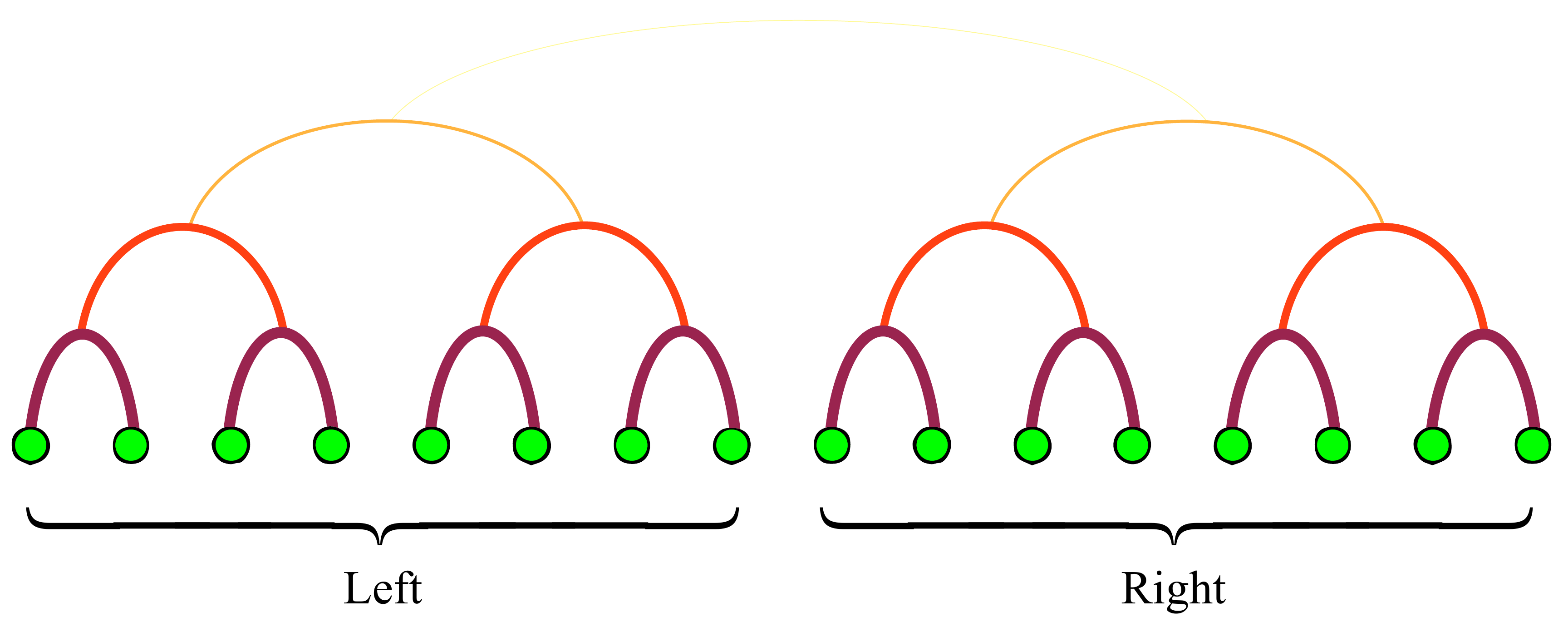}
\caption{\label{fig:esempio} Schematic representation of the hierarchical topology, that underlies the system under study: green spots represent nodes where spins/neurons live, while different colors and thickness for the links mimic different intensities in their mutual interactions: the brighter and thinner the link, the smaller the related coupling.}
\end{center}
\end{figure}
Before proceeding with the description of the Hopfield model, it is very instructive to make a step beside and revisit the ferromagnetic system described by the Curie-Weiss Hamiltonian (eq. $2$). The order parameter for the latter is given by the magnetization $m(S)$ defined as
\be
m(S)=\frac{1}{N}\sum_{i=1}^N S_i,
\ee
which, indeed, can distinguish between a paramagnetic/disordered phase ($m=0$) and a ferromagnetic phase characterized by spontaneous magnetization ($m \neq 0$). Moreover, we can write eq.($2$) also in terms of $m$ as
\be \label{eq:cw2}
H_{\textrm{Curie-Weiss}}(S) = - \frac{1}{N} \sum_{i<j}^{N,N} S_i S_j \sim -\frac{N}{2}m^2,
\ee
where a sub-leading term $\sum_i (S_i)^2/(2N) = 1/2$ has been neglected and we set $J=1$.
\newline
Restricting ourselves to the zero noise limit (for simplicity as entropy maximization can be discarded), following the minimum energy principle we see that the system tends to rearrange in such a way that $|m| \to 1$, corresponding to the configurations $\bold{S}=(+1,+1,...,+1)$ or $\bold{S}=(-1,-1,...,-1)$. If we read such a state as a neural configuration we would have a pathological state corresponding to {\em all} spins firing or quiescent. This point can be easily overcome by introducing the so-called Mattis gauge, namely by replacing $S_i \to \xi_i^{1}S_i$, where the set $\{ \xi^{1} \}$ may be drawn e.g., according to ($3$).
Via the Mattis gauge the Hamiltonian ($2$) can be rewritten as
\be
H_{\textrm{Mattis}} (S|\xi)= -\sum_{i<j}^{N,N} \xi_i^{1}\xi_j^{1} S_i S_j = -\frac{N}{2}m_1^2,
\ee
where $m_1$ is the Mattis magnetization defined as
\be
m_1=\frac1N \sum_{i=1}^N \xi_i^{1}S_i.
\ee
Reasoning exactly as before, in the low noise limit, following thermodynamic prescriptions, the system relaxes to the state with $|m_1| \to 1$, corresponding to a spin configuration $\bold{S}$ parallel (or anti-parallel) to the pattern $\bold{\xi}_1$. The relaxation to such a minimum (which now for a Shannon-McMillan argument is also the most likely and has only, on average, one half of the neurons firing) is seen as the {\em retrieval} of the (unique) stored pattern encoded by the string $\xi^1$. \\Now, enhancing the network capability, in such a way that the stored patterns are $p>1$, requires to abandon the ferromagnetic context as the system must be able to develop several free energy minima, each corresponding to the retrieval of a different pattern. This passage is formally straightforward: one simply introduces a sum over the patterns labelled as $\mu=1,...,p$ in the Mattis Hamiltonian, thus obtaining the Hopfield Hamiltonian ($4$).
\newline
When $p$ is large, that is comparable with the system size (thus in the so-called high-storage regime where $p$ scales as $N$, $p=\alpha N$ with $\alpha \in \mathbb{R}^+$), and as $N$ approaches infinity (as we deal with the so-called {\em thermodynamic limit} $N \to \infty$), for $\alpha > \alpha_c \sim 0.14$, retrieval properties are lost (and, for $p \to \infty$ quicker than $N$ the Hebbian coupling approaches a standard Gaussian $\mathcal{N}[0,1]$), hence the model collapses to the Sherrington-Kirkpatrick model for spin-glasses \cite{amit,howglassyNN}. In this regime neural capabilities are lost due the presence of too much disorder that splits the phase space into an amount of minima that scales exponentially with the system size \cite{MPV}. In the present paper we will work away from this {\em black out} limit focusing on the low storage scenario, where $p$ is either finite or growing much slower than $N$ (e.g. logarithmical), in such a way that $\lim_{N \to \infty} (p/N) \to 0$.

As mentioned above, as long as the noise is low enough, the system can relax in a (free) energy minimum: for the Hopfield model decribed by (\ref{HMF}) there exist overall $2p$ absolute minima corresponding to the configurations $S_i = \xi_i^{\mu}$ for all $i=1,...,N$; each minima encodes for the retrieval of a different pattern and the factor $2$ accounts for gauge symmetry $S_i \to -S_i$. The relaxation to the minimum corresponding to the, say, $k$-th pattern is evidenced by $m_k \neq 0$ and $m_i =0, \forall i\neq k$ (the latter holding on the average as patterns $\xi$'s are orthogonal -in the thermodynamic limit-).
The particular minimum selected depends on the external field (if present) and on the initial state of the system.
\newline
We stress that, since each pattern is built of by $N$ bits of information $\xi_i^{\mu} = \pm 1$, its retrieval involves the coordination of the whole network and the system can only retrieve patterns singularly, that is, one pattern at a time. For this reason this kind of processing is referred to as \emph{serial}.
\newline
This feature can be overcome and the neural network made able to perform \emph{parallel} retrieval, thus giving rise to the so called {\em multitasking associative network}  \cite{PRL}, by allowing for blank entries in the Hebbian kernel, that is, pattern entries are extracted once for all from
\be \label{eq:xi_blank}
P(\xi_i^{\mu}) = \left[\frac{1+a}{2} \delta(\xi_i^{\mu}-1) \frac{1+a}{2}\delta(\xi_i^{\mu}+1)\right] + a \delta(\xi_i^{\mu}),
\ee
where $a \in [-1,+1]$ tunes the amount of null-entries in the bit-strings.
\newline
Let us try to infer the effects of ($9$) on the retrieval process by focusing for simplicity on a simple case with $N=8$ and two toy-patterns $\bold{\xi}^{1}= (+1,+1,+1,+1,0,\\0,0,0)$ and $\bold{\xi}^{2}=(0,0,0,0,-1,-1,-1,-1)$, and with the external field (the stimuli) pointing to the first minimum. In suitable regions of phase space (where the network retrieves), the system will try to align with the first pattern, such that the first four neurons will be all firing. The remaining neurons do not receive any information from the pattern $\xi^1$, nevertheless, as the Hopfield Hamiltonian is a quadratic form in the Mattis magnetizations, (free)-energy minimization is better achieved if the remaining neurons align with the second pattern (instead of random reshuffling), such that the final state will be $\bold{S}=(+1,+1,+1,+1,-1,-1,-1,-1)$, and we say that the system has spontaneously perfectly retrieved the two patterns. An analogous behavior emerges for arbitrary $p$ patterns: the system tends to relax to a state where the Mattis magnetizations related to a subset of patterns are strictly non zero.
The performance of this network crucially depends on how $a$ is tuned as analyzed in details in \cite{ton1,ton2,Dantoni,PRE} for the low-storage and the high-storage regimes, respectively.

\subsection{The neural network on a hierarchical topology}
We now start our investigation of a neural network embedded in the hierarchical topology depicted in Fig.~$1$.
As mentioned, two main difficulties are interplaying: the complexity of the emergent energy landscape (essentially due to frustration in the coupling pattern) and the non-mean-field nature of the model (essentially due to the inhomogeneity of the network architecture).
It is therefore safer to proceed by steps discussing first the hierarchical ferromagnet (hence retaining only the second difficulty), known as Dyson hierarchical model (DHM). Then, via the Mattis gauge we reach a Mattis hierarchical model (MHN) and finally we extend to the Hopfield hierarchical model (HHM).

%



The Dyson hierarchical model \cite{dyson} is a system made of $N$ binary (Ising) spins $S_i = \pm 1$, $i=1,...,N$ in mutual interaction and built recursively in such a way that the system at the $(k+1)$-th iteration contains $N=2^{k+1}$ spins and is obtained by taking two replicas of the system at the $k$-th iteration (each made of $2^k$ spins) and connecting all possible couples with overall $\binom{N}{2}$ couplings equal to $- J / 2^{\sigma (k+1)}$, $J$ and $\sigma$ being real scalars tuning the interaction strength: the former acts uniformly over the network, the latter triggers the decay with the ``distance'' among spins. The resulting Hamiltonian can be written recursively as
\be\label{dyson}
H_{k+1}^{\textrm{Dyson}}(S|J,\sigma)=  H_{k}^{\textrm{Dyson}}(\mathbf{S}_1|J,\sigma) + H_{k}^{\textrm{Dyson}}(\mathbf{S}_2|J,\sigma) - \frac{J}{2^{2\sigma(k+1)}}\sum_{i<j}^{2^{k+1}}S_i S_j,
\ee
where $\bold{S}_1  = \{ S_i \}_{i=1}^{2^k}$ and  $\bold{S}_2  = \{ S_j \}_{i=2^k+1}^{2^{k+1}}$, while $H_0^{\textrm{Dyson}}\equiv 0$.

Before proceeding it is worth stressing that the parameters $J$ and $\sigma$ are bounded as $J>0$ and $\sigma \in (\frac{1}{2}, \ 1)$: the former trivially arises from the ferromagnetic nature of the model which makes neighboring spin to ``imitate'' each other, while the latter can be understood by noticing that for $\sigma >1$ the interaction energy goes to zero in the thermodynamic limit\footnote{The sum $\sum_{i<j}^{2^{k+1}}$ brings a contribution scaling like $2^{2(k+1)} \sim N^2$, while the pre-factor scales as $2^{-2 \sigma(k+1)} \sim N^{-2\sigma}$, thus, when
$\sigma>1$ the internal energy (the thermodynamical expectation of the Hamiltonian normalized over the system size) is overall vanishing in the thermodynamic limit $k \to \infty$.}, while for $\sigma < \frac{1}{2}$ the interaction energy is no longer linearly-additive implying thermodynamic instability\footnote{The sum $\sum_{i<j}^{2^{k+1}}$ brings a contribution scaling like $2^{2(k+1)} \sim N^2$, while the pre-factor scales as $2^{-2 \sigma(k+1)} \sim N^{-2\sigma}$, thus, when
$\sigma<\frac{1}{2}$ the intensive energy is overall divergent in the thermodynamic limit $k \to \infty$.}.
Moreover, this model is intrinsically {\em non-mean-field} because a notion of metrics, or distance, has been implicitly introduced: two nodes are said to be at distance $d$ if they get first connected at the $d$-th iteration. In general, calling $d_{ij}$ the {\em distance} between the spins $i, \ j$, (thus $d_{ij}=1,...,k+1$), we can associate to each couple a distant-dependent coupling $J_{ij}$ and rewrite ($10$) in a more familiar form as
\be\label{gallo}
H_{k+1}^{\textrm{Dyson}}(S|J,\sigma)= - \sum_{i<j}J_{ij}S_i S_j,
\ee
where
\be \label{eq:Jdef}
J_{ij}=\sum_{l=d_{ij}}^{k+1} \frac{J}{2^{2\sigma l}}= J\frac{4^{\sigma-d_{ij}\sigma}-4^{-k \sigma-\sigma}}{4^{\sigma}-1}.
\ee
The next step is to gauge the spins \textit{\`a} la Mattis, namely, once extracted quenched values for the pattern entries $(\xi_i^{\mu})_{\mu=1}$ from the distribution
\be
P(\xi_i^{\mu})=\frac12 \delta(\xi_i^{\mu}-1) + \frac12 \delta(\xi_i^{\mu}+1),
\ee
we replace $S_i$ with $\xi^1 S_i$. This results in the following hierarchical Mattis model
\be\label{mattis}
H_{k+1}^{\textrm{Mattis}}(S|J,\sigma)= - \sum_{i<j}J_{ij}\xi_i^1 \xi_j^1 S_i S_j.
\ee
\newline
Finally, summing over $p$ patterns, we obtain the Hopfield hierarchical model (HHM) that reads as (for $ J=1$)
\begin{eqnarray}\label{HHM}
\nonumber
H_{k+1}^{\textrm{Hopfield}}(S|\xi,\sigma) &=& H_{k}^{\textrm{Hopfield}}(S_1|\xi,\sigma) + H_{k}^{\textrm{Hopfield}}(S_2|\xi,\sigma) \\
&-& \frac12 \frac{1}{2^{2\sigma(k+1)}}\sum_{\mu=1}^p\sum_{i,j=1}^{2^{k+1}}\xi_i^{\mu}\xi_j^{\mu}S_i S_j,
\end{eqnarray}
with $H_0^{\textrm{Hopfield}}\equiv 0$ and $\sigma$ still within the previous bounds, i.e. $\sigma \in (\frac{1}{2}, 1)$.
As anticipated, here we restrict the analysis to low storage limit only: recalling $N=2^{k+1}$, we can fix $p $ finite at first so to move straightforwardly  from the DHM to the HHM (as the notion of distance is preserved) and, posing
\be
J_{ij} = \frac{4^{\sigma-d_{ij}\sigma}-4^{-k\sigma-\sigma}}{4^{\sigma}-1}\sum_{\mu=1}^p \xi_i^{\mu}\xi_j^{\mu},
\ee
we can write equivalently the Hamiltonian ($15$) in the more compact form
\be
H_{k+1}^{\textrm{Hopfield}}(S|\xi,\sigma) = - \sum_{i<j}^{2^{k+1}}J_{ij}S_i S_j.
\ee
Thus in the HHM the Hebbian prescription is coupled with (or ''weighted by'' \cite{jtb2,tirozzi}) a function of the neuron's distance.

In the following, in order to analyze in depth the system performance and the properties of hierarchical retrieval, we tackle the problem from different perspectives, each developed in a dedicated section. In particular, the next setion is devoted to the statistical mechanical route, fo which we report only results (as the methodologies underlying such achievements are still extremely technical and have been presented to the pertinent Community \cite{next}). As through this path a full analytical solution still lacks, further investigations must be addressed: indeed in Sec.$3$ we largely exploit outcomes from signal-to-noise studies, while numerical simulations are presented in section $4$.

\section{Insights from statistical mechanics} \label{sec:SM}

Here we summarize findings that can be achieved by suitably extending interpolation techniques \cite{broken,hightempGT} beyond the mean-field paradigm: it is important to stress once more that, as this strand gives only (not-mean-field) bounds on the free energy (and not the full solution), the self-consistencies that result are not the true self-consistencies of the model, thus motivating the next Sections.

\subsection{Pure/Ferromagnetic and Parallel/Mixed free energies in the Dyson model}

As the Hamiltonian $H_{k+1}(S|J,\sigma)$ is given (see eq. \ref{dyson}) and the noise level $\beta^{-1}=T$ (where $T$ stands for {\em noise} for historical reasons) introduced, it is possible to define the partition function $Z_{k+1}(\beta,J,\sigma)$ at finite volume $k+1$ as
\be
Z_{k+1}(\beta,J,\sigma) = \sum_{\{ S \} } \exp\left[-\beta H_{k+1}(S|J,\sigma)\right],
\ee
and the related free energy $f_{k+1}(\beta,J,\sigma)$, namely the intensive logarithm of the partition function, as
\be
f_{k+1}(\beta,J,\sigma)=\frac{1}{2^{k+1}}\log\sum_{\{S\}}\exp \left [-\beta H_{k+1}(\vec{S})+h \sum_{i=1}^{2^{k+1}}S_i \right],
\ee
where the sum runs over all possible $2^{2^{k+1}}$ spin configurations. Note that the  usual free energy $\tilde{f}$ is related to $f$ by $\tilde{f}(\beta)=-\beta f(\beta)$, hence we will find thermodynamic equilibria checking the maxima of $f(\beta)$ and not the minima.
\newline
We are interested in an explicit expression of the infinite volume limit of the intensive free energy, defined as
\be
f(\beta,J,\sigma)= \lim_{k\to\infty}f_{k+1}(\beta,J,\sigma),
\ee
in terms of suitably introduced magnetizations $m$, that act as order parameters for the theory. In fact, as the free energy is just the difference between the internal energy $E$ of the system (i.e. the mean-value of the Hamiltonian) weighted by $\beta$, and the entropy $S$, namely $f(\beta,J,\sigma) = - \beta E(\beta,J,\sigma) + S (\beta,J,\sigma)$, extremization of the free-energy over the order parameters equals to imposing thermodynamic prescriptions (i.e. minimum energy and maximum entropy principles) and therefore allows us to get a description of the thermodynamic equilibria of the system in terms of the self-consistencies for these $m$'s.
\newline
To this task we introduce the global magnetization $m$, defined as the limit $m=\lim_{k\to\infty}m_{k+1}$ where
\be
m_{k+1}=\frac{1}{2^{k+1}}\sum_{i=1}^{2^{k+1}}S_i,
\ee
and, recursively and with a little abuse of notation, level by level (over $k$ levels)
the $k$ magnetizations $\vec{m}_a,...,\vec{m}_k$, as the same $k\to\infty$ limit of the following quantities (we write explicitly only the two upper magnetizations related to the two main clusters {\em left} and {\em right} -see Fig.$1$-):
\be
m_{k}^1=\frac{1}{2^{k}}\sum_{i=1}^{2^{k}}S_i, \ \ \ m_{k}^2=\frac{1}{2^{k}}\sum_{i=2^k+1}^{2^{k+1}}S_i,
\ee
and so on. The {\em thermodynamical averages} are denoted by the brackets $\langle \cdot \rangle$ such that, e.g. for the observable $m_{k+1}(\beta,J,\sigma)$, we can write
\be
\langle m_{k+1}(\beta,J,\sigma) \rangle = \frac{\sum_{\sigma} m_{k+1}e^{-\beta H_{k+1}(\vec{S}|J,\sigma)}}{Z_{k+1}(\beta,J,\sigma)},
\ee
and clearly $\langle m(\beta,J,\sigma) \rangle = \lim_{k \to \infty} \langle m_{k+1}(\beta,J,\sigma) \rangle$.
\newline
Starting with the pure ferromagnetic case, which mirrors here the serial retrieval of a single pattern in the Hopfield counterpart,  its free energy can be bounded as (see also \cite{castellana-jsp})
\be\small
f(h,\beta,J,\sigma)\geq  \sup_{m} \left\{\log 2+\log\cosh\Big[h+\beta mJ(C_{2\sigma-1}-C_{2\sigma})\Big]-\frac{\beta J}{2}(C_{2\sigma-1}-C_{2\sigma})m^2\right\},
\ee
where
\begin{eqnarray}
C_{2\sigma} &=& \frac{1}{2^{2\sigma}-1},\\
C_{2\sigma-1} &=&\frac{1}{2^{2\sigma+1}-1}.
\end{eqnarray}
Now, let us suppose that, instead of a global ordering, the system can be effectively split in two parts (the two largest communities called {\em left} and  {\em right} in Fig.$1$), with two different magnetizations $m_{left}=m_1$ and $m_{right}=m_2$; we also assume $m_{left}=-m_{right}$.  Through the interpolative route we approach a bound for the free energy related to such a mixed state. We stress the fact that the upper link, connecting the two communities with opposite magnetization, remains and it gives a contribute $m$ in the system as (see also \cite{next})
\begin{eqnarray}\nonumber\small\small
f_{k+1}&\geq& \frac{1}{2}\log\cosh\left\{h+\beta J \left[m (2^{(k+1)(1-2\sigma)})+ m_1 \left(\sum_{l=1}^{k}2^{l(1-2\sigma)}-\sum_{l=1}^{k+1}2^{-2l\sigma}\right)\right]\right\} \\ \nonumber &+&\frac{1}{2}\log\cosh \left\{h+\beta J \left[m(2^{(k+1)(1-2\sigma)})+ m_2 \left(\sum_{l=1}^{k}2^{l(1-2\sigma)}-\sum_{l=1}^{k+1}2^{-2l\sigma}\right)\right] \right\} \\&-&\frac{\beta J}{2}\left[\left(\sum_{l=1}^{k}2^{l(1-2\sigma)}-\sum_{l=1}^{k+1}2^{-2l\sigma}\right)\left(\frac{m_1^2+m_2^2}{2}\right)-2^{(k+1)(1-2\sigma)}m^2\right]\nonumber\\&+& \log 2.
\end{eqnarray}
Notice that, thanks to the gauge simmetry $S_i\rightarrow -S_i$, the state considered mirrors the parallel retrieval of two patterns in the Hopfield counterpart.
%
%
Identifying $m_1=m_2=m$ we recover the previous bound as expected, and, quite remarkably, in the thermodynamic limit the two free energies assume the same values, thus serial and parallel retrieval are both equally accomplished by the network.
Imposing thermodynamic stability we obtain the following self-consistencies
\be\label{scD}
m_{1,2}=\tanh(h+\beta Jm_{1,2}(C_{2\sigma-1}-C_{2\sigma})),
\ee
whose behavior is depicted in Fig. 2.

\subsection{Serial versus parallel retrieval in Hopfield hierarchical model}

Guided by the ferromagnetic model just described, we now turn to the hierarchical Hopfield model (HHM) and start its analysis from a statistical mechanical perspective, namely we infer the thermodynamic behavior of a system described by the following recursive Hamiltonian
\begin{eqnarray}
H_{k+1}^{HHM}(S|\xi,\sigma)&=& H_{k}^{HHM}(S_1|\xi,\sigma) +H_{k}^{HHM}(S_2|\xi,\sigma) \\ \nonumber
&-& \frac12 \frac{1}{2^{2\sigma(k+1)}}\sum_{\mu=1}^p \sum_{i,j}^{2^{k+1}}\xi_i^{\mu}\xi_j^{\mu}\sigma_i\sigma_j.
\end{eqnarray}
To this task, we introduce suitably $p$ Mattis magnetizations (or Mattis overlaps), over the whole system, as
\be
m^{\mu} = \frac{1}{2^{k+1}}\sum_{i=1}^{2^{k+1}}\xi_i^{\mu} S_i, \text{ }\mu\in[1,p].
\ee
Even in this context, the definition above can account for the state of inner clusters by the sum over the (pertinent) spins. For instance, focusing on the two larger communities we have the $2p$ Mattis magnetizations
\be
m^{\mu}_{left}=\frac{1}{2^k}\sum_{i=1}^{2^k}\xi_i^{\mu} S_i, \ \ \ \ %
m^{\mu}_{right}=\frac{1}{2^k}\sum_{i=2^k+1}^{2^{k+1}}\xi_i^{\mu} S_i,
\ee
with  $\mu\in[1,p]$. Again, we will not enter in the mathematical details concerning non-mean-field bounds for the model free energy (as they can be found in \cite{next}), while we streamline directly the physical results.
\newline
Still mirroring the previous section, we are interested in obtaining a bound limiting the free energy of the HHM, the latter being defined as the $k \to \infty$ limit of $f_{k+1}$, whose expression reads
\be
f_{k+1}(\beta,\{h_{\mu}\},\sigma)=\frac{1}{2^{k+1}}\log\sum_{\{S\}}\exp \left [-\beta H_{k+1}(\vec{S})+ \sum_{\mu=1}^p h^{\mu} \sum_{i=1}^{2^{k+1}}S_i \right],
\ee
where we accounted also for $p$ external stimuli $h^{\mu}$.
\newline
The non-mean field bound for serial processing free energy reads as
\ba
\nonumber
f(\beta, \{h^{\mu}\},p) &\geq& \sup_{m} [\log 2+\Big\langle\log\cosh\Big(\sum_{\mu=1}^{p} \Big[h^{\mu}+\beta m^{\mu}(C_{2\sigma-1}-C_{2\sigma})\Big]\xi^{\mu}\Big)\Big\rangle_{\xi}\\
&-&\frac{\beta}{2}\sum_{\mu=1}^p \langle (m^{\mu})^2 \rangle_{\xi}(C_{2\sigma-1}-C_{2\sigma})],
\ea
with optimal order parameters fulfilling
$$
\langle m^{\mu} \rangle_{\xi}=\langle \xi^{\mu}\tanh [\beta\sum_{\nu=1}^p \left[ h^{\nu}+(C_{2\sigma-1}-C_{2\sigma})m^{\nu} \right] \xi^{\nu}]\rangle_{\xi},
$$
and whose critical noise is $\beta^{NMF}_c=C_{2\sigma-1}-C_{2\sigma}$, where the index $NMF$ stresses that the estimate was obtained through a non mean field bound of the free energy.
\newline
Of course we can assume again that the two different families of Mattis magnetizations $(\{m^{\mu}_{1,2}\}_{\mu=1}^p)$ (those playing for the two inner blocks of spins {\em left} and {\em right} lying under the $k+1$-th level) behave independently as the higher links connecting them go to zero quickly for $k \to \infty$ and we can start the interpolative machine: following this way we generalize the serial processing analysis to a two-pattern parallel retrieval analysis, which results in the following bound for the related free energy:
\begin{eqnarray}
\nonumber
&& f(\beta, \{h_{\mu}\},p)\geq\sup_{\{m^{\mu}_{1,2}\}} [\log 2+\frac 1 2 \Big\langle \log\cosh \Big\{ \sum_{\mu=1}^{p} \Big[ h^{\mu}+\beta m^{\mu}_1 \Big(\sum_{l=1}^{k}2^{l(1-2\sigma)} \\
\nonumber
&& -\sum_{l=1}^{k}2^{l(-2\sigma)} \Big)  + \beta m^{\mu}2^{(k+1)(1-2\sigma)}\Big]  \xi^{\mu} \Big\}\Big\rangle_{\xi} + \frac{1}{ 2} \Big\langle\log\cosh  \Big\{ \sum_{\mu=1}^{p}  \Big[h^{\mu} + \beta m^{\mu}_2\\
\nonumber
&& \times [\sum_{l=1}^{k}2^{l(1-2\sigma)}-\sum_{l=1}^{k}2^{l(-2\sigma)}] + \beta m^{\mu}2^{(k+1)(1-2\sigma)}\Big] \xi^{\mu} \Big\} \Big\rangle_{\xi} -\frac{\beta}{2}\Big[\sum_{l=1}^{k}2^{l(1-2\sigma)} \\
\nonumber
&& -\sum_{l=1}^{k}2^{l(-2\sigma)}\Big] \cdot \sum_{\mu=1}^p \frac{\langle{(m^{\mu}_1)^2}\rangle_{\xi}+{\langle (m^{\mu}_2)^2\rangle_{\xi}}^2}{2} -\frac{\beta}{2}2^{(k+1)(1-2\sigma)}\sum_{\mu=1}^p \langle (m^{\mu})^2 \rangle_{\xi},
\end{eqnarray}
Here we do not investigate further the parallel retrieval of larger ensembles of patterns, as the way to proceed is identical to the outlined one, but we simply notice that, if we want the system to handle $M$ patterns, hence we assume it effectively splits $M$ times into sub-clusters until the $k+1-M$ level, then the procedure keeps on working as long as
\be
\lim_{k\to\infty} \sum_{l=k+1-M}^{k+1} 2^{l(1-2\sigma)} \sum_{\mu=1}^p m^{\mu}_{l}=0.
\ee
Since the magnetizations are bounded, in the worst case we have
\begin{eqnarray}
\sum_{l=k+1-M}^{k+1} 2^{l(1-2\sigma)} \sum_{\mu=1}^p m^{\mu}_{l}&\leq& p \sum_{l=k+1-M}^{k+1} 2^{l(1-2\sigma)} \nonumber\\
&\leq& p \sum_{l=k+1-M}^\infty 2^{l(1-2\sigma)} \propto  2^{(1-2\sigma)(k+1-M)} p.
\end{eqnarray}
If we want the system to handle up to  $p$ patterns, we need $p$ different blocks of spins and then $M=\log(p)$.

\begin{figure}[tbh!] \begin{center}
\includegraphics[width=.9\textwidth]{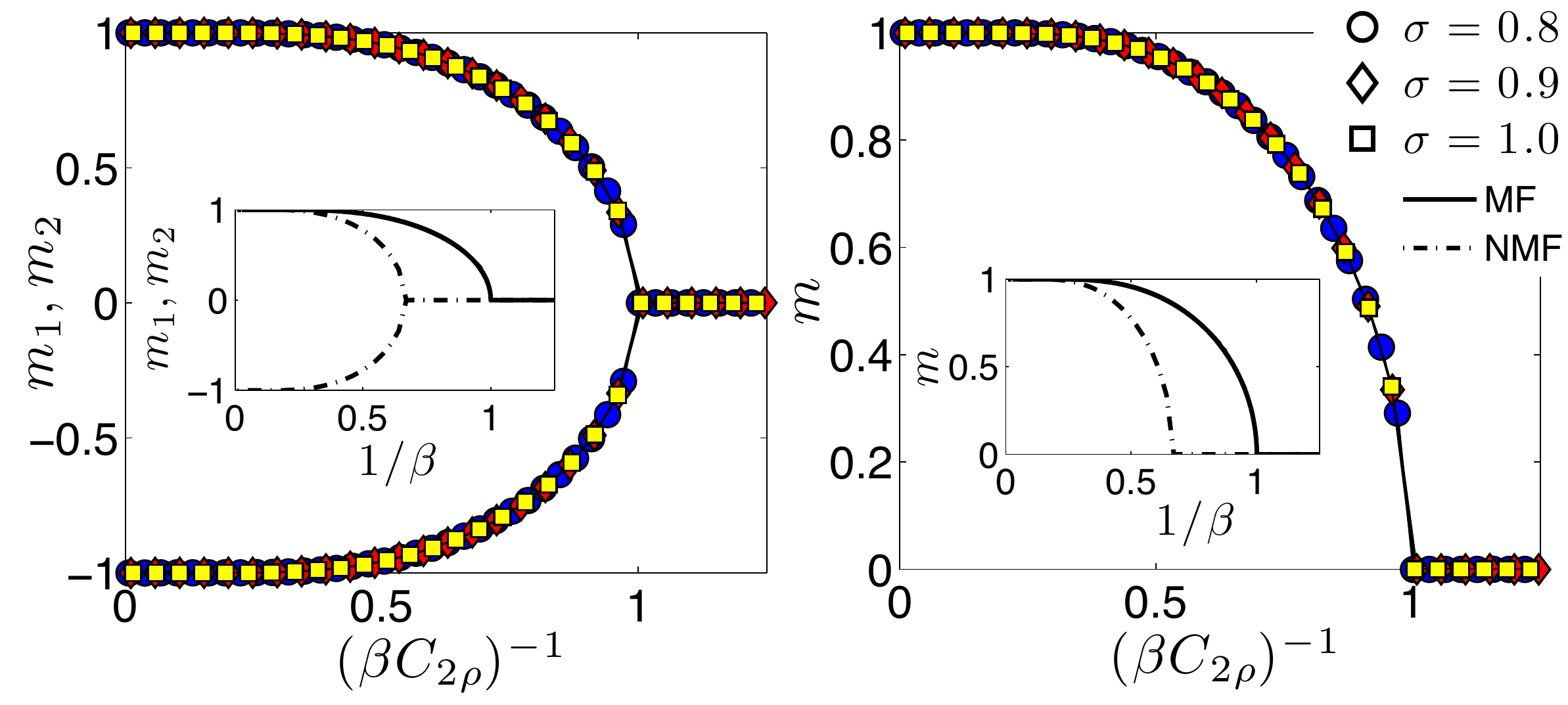}
\caption{\label{fig:esempio} Main plots: numerical solutions of the non-mean-field self-consistent equations for the parallel state (left panel) and for the pure state (right panel) of the Dyson model (see Eq. \ref{scD}) obtained for different values of $\sigma$ (as explained by the legend) and plotted versus a rescaled noise. Note that by rescaling the noise the dependence on $\sigma$ is lost and all curves are collapsed.
Insets: comparison between the numerical solutions of the non-mean-field self-consistent equations (dashed line) and of the mean-field self-consistent equations (solid line) as a function of the noise and for fixed $\sigma=1$ (see Eq.~\ref{scD}).
Notice that for the Hopfield hierarchical model, numerical solutions for the Mattis magnetizations pertaining to the pure and to the mixed states are the same.}
\end{center}
\end{figure}

\section{Insights from signal-to-noise techniques}

Results from statistical mechanics gave stringent hints on the network's behavior, however they act as bounds only. \newline
This requires further inspection via other techniques: the first route we exploit is signal-to-noise. Through the latter, beyond generally confirming the predictions obtained via the first path, we obtain sharper statements regarding the evolution of the Mattis order parameters.
These two approaches are complementary: while statistical mechanics describes the system with $N\rightarrow\infty$ and $\beta<\infty$, with the signal-to-noise technique we inspect  the regime $N<\infty$ and $\beta\rightarrow\infty$.

\subsection{A glance at the fields in the Dyson network}

Plan of this Section is to look at the dynamically stable configurations of the neurons, that is to say, we investigate the configurations (global and local minima) that imply each neuron $S_i$ to be aligned with its corresponding field $h_i[S]$, i.e.  $S_i \, h_i[S] >0, \forall i$. This approach basically corresponds to a negligible-noise statistical mechanical analysis but it is mathematically much more tractable.
\newline
We can rearrange the Dyson Hamiltonian in a useful form for such an investigation as follows
\be
H_{k+1}^{\text{Dyson}}(\{S_1...S_{2^{k+1}}\})=-\frac{J}{2}\sum_{\mu=1}^{k+1} \sum_{i=1}^{2^{k+1}} S_i  \left [ \sum_{l=\mu}^{k+1} \left( \frac{1}{2^{2\sigma}} \right)^l \right] \sum_{\{j\}:d_{ij}=\mu} S_j,
\ee
thus, highlighting the field $h_i$ insisting on the spin $S_i$ we can write
\begin{eqnarray}\label{eq:HDyson}
H_{k+1}^{\text{Dyson}}[\{S_1...S_{2^{k+1}}\}]&=&- \sum_{i=1}^{2^{k+1}} S_i h_i[\mathbf{S}],  \\
h_i[\mathbf{S}]&=&J\sum_{\mu=1}^{k+1} \left[ \sum_{l=\mu}^{k+1} \left(\frac{1}{2^{2\sigma}} \right)^l \right] \sum_{\{j\}:d_{ij}=\mu} S_j.
\end{eqnarray}

While Glauber dynamics will be discussed in Sec.~$4$ (dedicated to numerics), we just notice here that the microscopic law governing the evolution of the system can be defined as a stochastic alignment to local field $h_i[\mathbf{S}]$.
$$
S_i(t+\delta t)=\sign\left \{ \tanh\left[\beta h_i\left[ \mathbf{S}\left(t\right)\right)\right ]+\eta_i(t)\right \},
$$
where the stochasticity lies in the independent random numbers $\eta_i(t)$, uniformly distributed over the interval $[-1, 1]$ and tuned by $\beta$. The latter continues to rule the noise level even dynamically as it amplifies, or suppresses, the smoothness of the hyperbolic tangent; in particular, in the noiseless limit $\beta \rightarrow \infty$ we get
\be
S_i(t+\delta t)=\sign\left[h_i\left(\mathbf{S}(t)\right)\right].
\ee
This is crucial for checking the stability of a state as, if $S_i h_i[\mathbf{S}] >0$ $\forall \ i \ \in \ [1,N]$, the configuration $\{ \mathbf{S}\}$ is dynamically stable (at least for $\beta \to \infty$, as in the presence of noise there is a $\beta$-dependent probability to fluctuate away).

We keep the previous ensemble of non-independent order parameters $m_i^n$ defined in detail as
\be \label{eq:4}
m_i^n[\mathbf{S}]=\frac{1}{2^n} \sum_{j= 2^n\times i-(2^n-1)}^{2^n\times i}S_j \ \ \ \mbox{with } i=1,2,...,2^{k+1-n} \ \ \ \mbox{and } n=0,1,2,...k+1,
\ee
namely
$$
\begin{cases}
m_i^0=S_i & \mbox{with } i=1,2,..,2^{k+1},  \\
m_i^1=\frac{1}{2}\sum_{j=2i-1}^{2i}S_j & \mbox{with } i=1,2,..,2^{k} \ \to m_1^1=\frac{1}{2}\sum_{j=1}^{2}S_j, \\
m_i^2=\frac{1}{2^2}\sum_{j=2^2i-(2^2-1)}^{2^2i}S_j & \mbox{with } i=1,2,..,2^{k-1} \ \to m_1^2=\frac{1}{4}\sum_{j=1}^{4}S_j, \\
..... \\
m_1^{k+1}=\frac{1}{2^{k+1}}\sum_{j=1}^{2^{k+1}}S_j.
\end{cases}
$$

From Eq.~$38$, we get the following fundamental expression for the fields
\be
h_i[\mathbf{S}]= \left[ J \sum_{\mu=1}^{k+1} \left( \sum_{l=\mu}^{k+1}\frac{1}{2^{2\sigma}}\right)^l \right]2^{\mu-1}m_{f(\mu,i)}^{\mu-1},
\ee
where we used the relation  $m_{f(\mu,i)}^{\mu-1}=\sum_{\{j\}:d_{ij=\mu}}S_j$.
Thus the order parameters $m_{f(\mu,i)}^{\mu-1}$ represent the magnetizations assumed by spins that lie at distance $\mu$ from $S_i$. Note that the function $f(\mu,i)$ can be estimated through the floor function $\lfloor\cdot\rfloor$ (e.g., $\lfloor 3.14\rfloor=3$) as
$$
f(\mu,i)=\Big\lfloor \frac{i+(2^{\mu-1}-1)}{2^{\mu-1}}\Big\rfloor+ (-1)^{(\lfloor \frac{i+(2^{\mu-1}-1)}{2^{\mu-1}}\rfloor+1)}.
$$
Finally, we notice that the largest value allowed for a field -away from the boundary value $\sigma = 1/2$- for large $k$ approaches a plateau (whose boundaries -in the  $(k,\sigma)$ plane- are important for finite-size-scaling during numerical analysis), hence we can easily check the right field normalization
\begin{eqnarray}
Q(\sigma,k+1)=\sum_{\mu=1}^{k+1} J(\mu,k+1,\sigma)2^{\mu-1}=\nonumber\\=J\frac{2^{-2(k+1) \sigma } \left(2^{2 (k+2) \sigma }-2^{k+2 \sigma +2}+2^{k+2}+4^{\sigma }-2\right)}{-3 \times 4^{\sigma }+16^{\sigma }+2},
\end{eqnarray}
as $Q(\sigma,k)$ represents the largest value allowed by a field.
\newline
Note that in the thermodynamic limit
\be
\lim_{k\to\infty}Q(\sigma,k)=Q(\sigma)=J\frac{2^{2 \sigma }}{-3 \times 4^{\sigma }+4^{2 \sigma }+2},
\ee
that is $Q$ is always bounded whenever $\sigma>\frac{1}{2}$.

\subsection{Metastabilities in the Dyson network: Noiseless case.}

We can now proceed to the stability analysis explaining in details a few test cases that show how to proceed for any other case of further interest:
\newline
$[a]$ the global ferromagnetic state, i.e. $S_i = +1$, $i \in (1,...,2^{k+1})$.
\newline
$[b]$ the parallel/mixed state, i.e. the first half of spins up and the second half down, thus $S_i = +1$, $i \in (1,...,2^{k})$ and $S_i = -1$, $i \in (2^{k}+1,...,2^{k+1})$.
\newline
$[c]$ the dimer, i.e. $S_1=S_2=+1$ while $S_i = -1$ for all $i \neq (1,2)$.
\newline
$[d]$ the square, i.e. $S_1=S_2=S_3=S_4=+1$ while $S_i = -1$ for all $i>4$.

\quad

Let us go through each case analysis separately:
\begin{itemize}
\item  $[a]$  The global ferromagnetic state $S_i=+1 \ \ \forall i \in[1,2^{k+1}] \ \  \Rightarrow \ \ m_i^n[\mathbf{S}]=1 \ \forall i,n $ has fields
\begin{eqnarray}
&\Rightarrow& h_i[\mathbf{S}]=J\frac{4^{-(k+1) \sigma } \left[2^{2 (k+2) \sigma }-2^{k+2+2 \sigma }+2^{k+2}+4^{\sigma }-2\right]}{-3 \times 4^{\sigma }+16^{\sigma }+2},\\ \nonumber
\ \ \\
&\Rightarrow& h_i[\mathbf{S}]>0 \ \forall k,\sigma\in (1/2,1).
\end{eqnarray}
Thus, the configuration $S_i=+1 \ \ \forall i \in[1,2^{k+1}]$ is stable in the noiseless limit  $ \forall\sigma \in[\frac{1}{2},1]$. In the thermodynamic limit $k\rightarrow\infty$ we have
$$
h_i[\mathbf{S}]=J \frac{4^{\sigma }}{-3 \times 4^{\sigma }+16^{\sigma }+2}.
$$
To address network's behaviour in the presence of noise, fixing $J=1$ without loss of generality, we can look at the solution of the following equation
\be
\tanh(\beta h_i[\mathbf{S}] )\simeq1 \Rightarrow \tanh \left( \beta \frac{4^{\sigma }}{-3 \times 4^{\sigma }+16^{\sigma}+2} \right)  \simeq 1.\label{crit1}
\ee
This allows to find the curve $\beta_{c}^{\textrm{no \ errors}}(\sigma)$ versus $\sigma$ (shown in Fig.$3$). In fact, we know  that, at the time $t+\delta t$, the system obeys the dynamics
$$S_i(t+\delta t)=\sign(\tanh(\beta h_i(\mathbf{S}))+\eta_i),$$
where $\eta_i$ is a random variable, whose value is uniformly distributed in $[-1,1]$. Imposing $\tanh(\beta h_i)\simeq 1$ we ask that $|h_i|\gg 1$, so the sign of the right hand side member of the equation is positive, thus the sign of $S_i$ at the time $t+\delta t$ is the same of the field $h_i$ at the time $t$.
Then, fixed $\sigma$, for every $\beta>\beta_{c}^{\textrm{no \ errors}}(\sigma)$ the state $S_i=+1 \ \ \forall i \in[1,2^{k+1}]$ is stable without errors.

\item $[b]$ The parallel/mixed state $S_j=+1 \ \ S_i=-1 \ \ \forall j \in[1,2^{k}]\ \ \forall i \in[2^{k}+1,2^{k+1}]$ has fields
\begin{eqnarray}
\nonumber
\Rightarrow h_j[\mathbf{S}]&=&J \frac{4^{-(k+1)\sigma } \left(2^{2 (k+2) \sigma }+2^{k+1+2 \sigma }-2^{k+1+4 \sigma} +4^{\sigma }-2\right)}{-3 \times 4^{\sigma }+16^{\sigma }+2}\\
&=&-h_i[\mathbf{S}]>0 \ \forall \ k+1 \geq 2,\\ \nonumber
 \ \\
&\Rightarrow& \lim_{k \to \infty} h_j[\mathbf{S}] = J \frac{1}{2^{1-2 \sigma }+4^{\sigma }-3},
\end{eqnarray}
thus the configuration $S_j=+1 \ \ S_i=-1 \ \ \forall j \in[1,2^{k}]\ \ \forall i \in[2^{k}+1,2^{k+1}] $ is stable in the noiseless limit $\forall \ k+1>2, \ \sigma\in (1/2,1)$.
Using the same arguments of the previous case, fixing $J=1$ without loss of generality, to infer network's behaviour in the presence of the noise we can look at the solution of the following equation
\be \label{eq:crit}
\tanh(\beta h_i[\mathbf{S}] )\simeq 1 \Rightarrow \tanh \left( \beta \frac{1}{2^{1-2 \sigma }+4^{\sigma }-3} \right) \simeq 1.
\ee
This allows to find the curve $\beta_{c}^{\textrm{no-errors}}(\sigma)$ versus $\sigma$ (see Fig.$3$).
Then, fixed $\sigma$, for every $\beta>\beta_{c}^{\textrm{no-errors}}(\sigma)$ the state $S_j=1 \ \ S_i=-1 \ \ \forall j \in[1,2^{k}]\ \ \forall i \in[1+2^{k},2^{k+1}] $ is stable without errors.
  So we can see how, in the thermodynamic limit, the state with all spins aligned $S_j=+1 \ \ \forall j \in[1,2^{k+1}]$ and the state with half spins pointing upwards and half pointing downwards  $S_j=+1 \ \ \forall j \in[1,2^{k}]\ \ S_i=-1 \ \ \ \forall i \in[1+2^{k},2^{k+1}] $ are both robust.
For an arbitrary finite value of $k$ it is possible to solve numerically  eq.~$49$ to get an estimate for $\beta_{c}^{\textrm{no-errors}}(\sigma)$ versus $\sigma$:  in Figure $3$ $\beta_{c}^{\textrm{no-errors}}(\sigma)$ is plotted for the state  $S_j=+1 \ \ S_i=-1 \ \ \forall j \in[1,2^{k}]\ \ \forall i \in[1+2^{k},2^{k+1}] $ and the state $S_i=+1 \ \ \forall i \in[1,2^{k+1}]$.


\item $[c]$ The dimer $S_j=+1 \ \ S_i=-1 \ \ \forall j \in[1,2]\ \ \forall i \in[3,2^{k+1}] $ has fields
\begin{eqnarray} \nonumber \small
&&h_1[\mathbf{S}]=h_2[\mathbf{S}]= \frac{2^{-2\sigma(k+1)}(2^{2\sigma(k+2)}+2^{k+2+2\sigma}-4^{1+(k+1)\sigma}-2^{k+2}-3\times 4^{\sigma}+6)}{(-3 \times 4^{\sigma }+16^{\sigma }+2)},\\  \nonumber \small
&&\lim_{k \to \infty}h_1[\mathbf{S}]=\lim_{k \to \infty}h_2[\mathbf{S}]= 2\cdot \frac{4^{\sigma }-4}{-3\times4^{\sigma }+16^{\sigma }+2}<0 \ \forall \sigma \in (1/2,1).
\end{eqnarray}
Therefore, the configuration  $S_j=+1 \ \ S_i=-1 \ \ \forall j \in[1,2]\ \ \forall i \in[3,2^{k+1}] $, in the thermodynamic limit, is unstable $\forall \ \sigma \in (1/2,1)$.


\item $[d]$
The square  $S_j=1 \ \ S_i=-1 \ \ \forall j \in[1,4]\ \ \forall i \in[5,2^{k+1}] $ has fields
\begin{eqnarray}
\nonumber \small
h_j[\mathbf{S},k]&=&-\frac{2^{1-2 (k+1) \sigma } \left(-2^{k+1+2 \sigma }+2^{2 k \sigma +1}+2^{k+1}+2^{2 \sigma +1}-4\right)}{-3 \times 4^{\sigma }+16^{\sigma }+2} \\
&-&\frac{-3 \times 4^{-({k+1}) \sigma }+2^{1-2 \sigma}+1}{1-4^{\sigma }},\\
\nonumber \small
h_j[\mathbf{S},k+1]&=&\frac{\left(2^{2 (k+3) \sigma }-2^{k+2+2 \sigma}+2^{k+2+4 \sigma}-2^{2 (k+1) \sigma +3}+7\times 2^{2 \sigma +1}-7 \times 16^{\sigma }\right)}{(-3\times 4^{\sigma }+16^{\sigma}+2)/(2^{-2 (k+2) \sigma })}
\end{eqnarray}
thus
$$
\lim_{k \to \infty} h_j[\mathbf{S}] =\frac{4^{-\sigma } \left(16^{\sigma }-8\right)}{-3 \times 4^{\sigma }+16^{\sigma }+2}=\begin{cases} >0, & \mbox{if }\sigma>\frac{3}{4} \\ <0, & \mbox{if } \sigma<\frac{3}{4}\end{cases}.
$$
Therefore, the configuration $S_j=+1 \ \ S_i=-1 \ \ \forall j \in[1,4]\ \ \forall i \in[5,2^{k+1}] $ in the limit $(k\rightarrow\infty)$ for $T=0$ is stable $\forall \ \sigma\in(\frac{3}{4},1)$
\end{itemize}

It is worth noticing that beyond the extensive meta-stable states (e.g. the parallel/mixed one) already suggested by the statistical mechanical route, stability analysis predicts that tighley connected modules (e.g. octangon, esadecagon, ...) with spins anti-aligned with respect to the bulk get dynamically stable in the thermodynamic limit: these {\em motifs} in turn are able to process small amount of information and an analysis of their capabilities can be found in \cite{ton1,ton2}, and their robusting is due to their intrinsic loopy structure.


\begin{figure}[!h!d] \label{fig2}
\centering
\includegraphics[width=9cm]{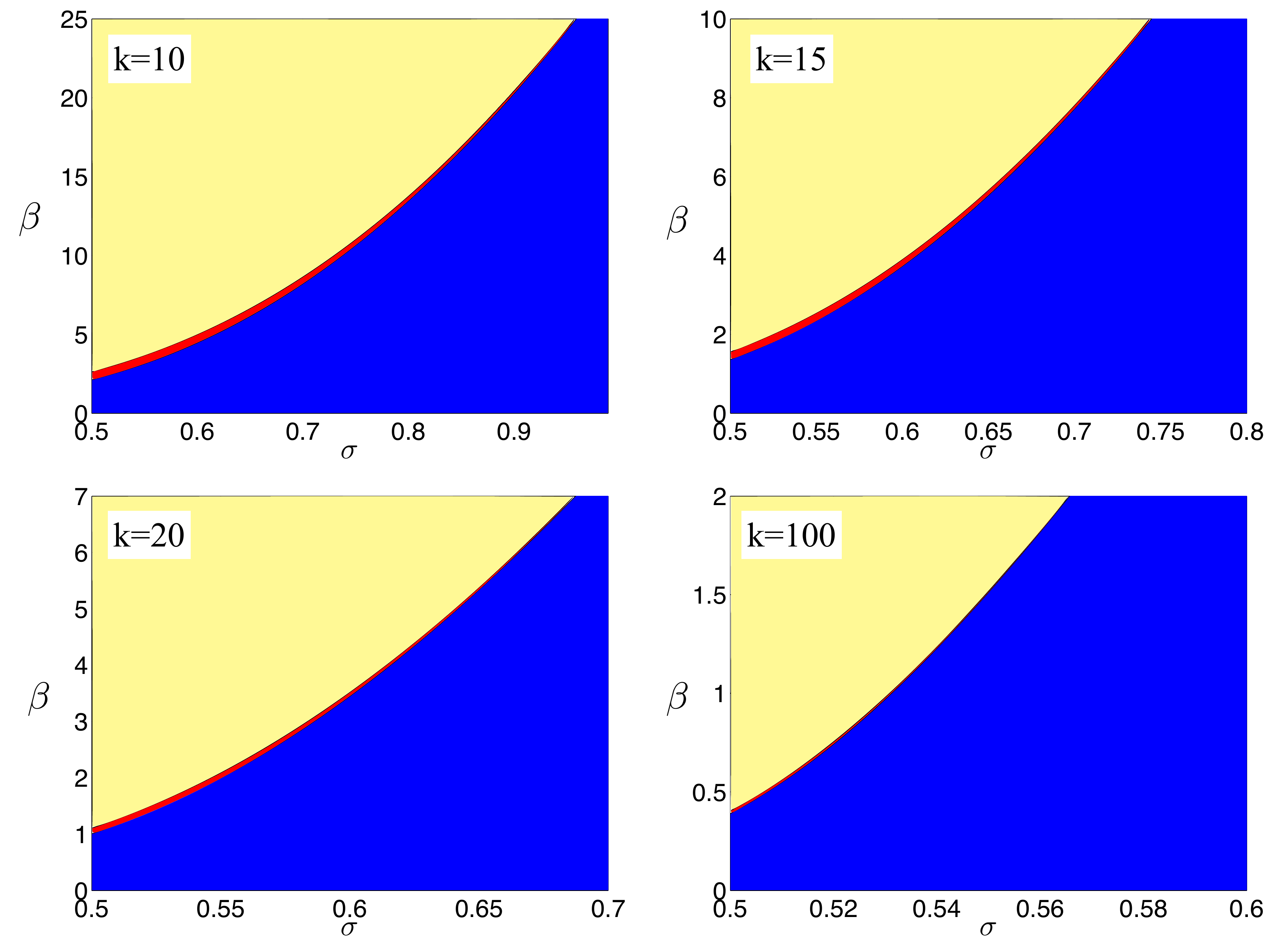}
\caption{Phase diagram for the perfect retrieval accomplished by a pure state ($S_i=+1$ $\forall i=1,...,2^{k+1}$)   and parallel state ($S_i=+1$ $\forall i=1,...,2^{k}$ and $S_i=-1$ $\forall i=2^{k}+1,...,2^{k+1}$). The line separating different regions  corresponds to numerical solution of $\beta_{c}^\textrm{no \ errors}[\sigma] \ $ versus $ \ \sigma$, obtained from $ (\ref{crit1})$ and  $(\ref{eq:crit})$ for different values of $k$ ($10, 15, 20, 100$ respectively). In yellow, the area where both the pure and parallel states are perfectly retrieved, while in blue the area where none of them is retrieved. The red line represents the area where only the pure state is stable: this region vanishes as $k$ gets larger (namely in the thermodynamic limit), hence confirming that the pure and the mixed state are both global minima.}
\end{figure}

\begin{figure}[!hd] \label{fig6}
\centering
\includegraphics[width=9cm]{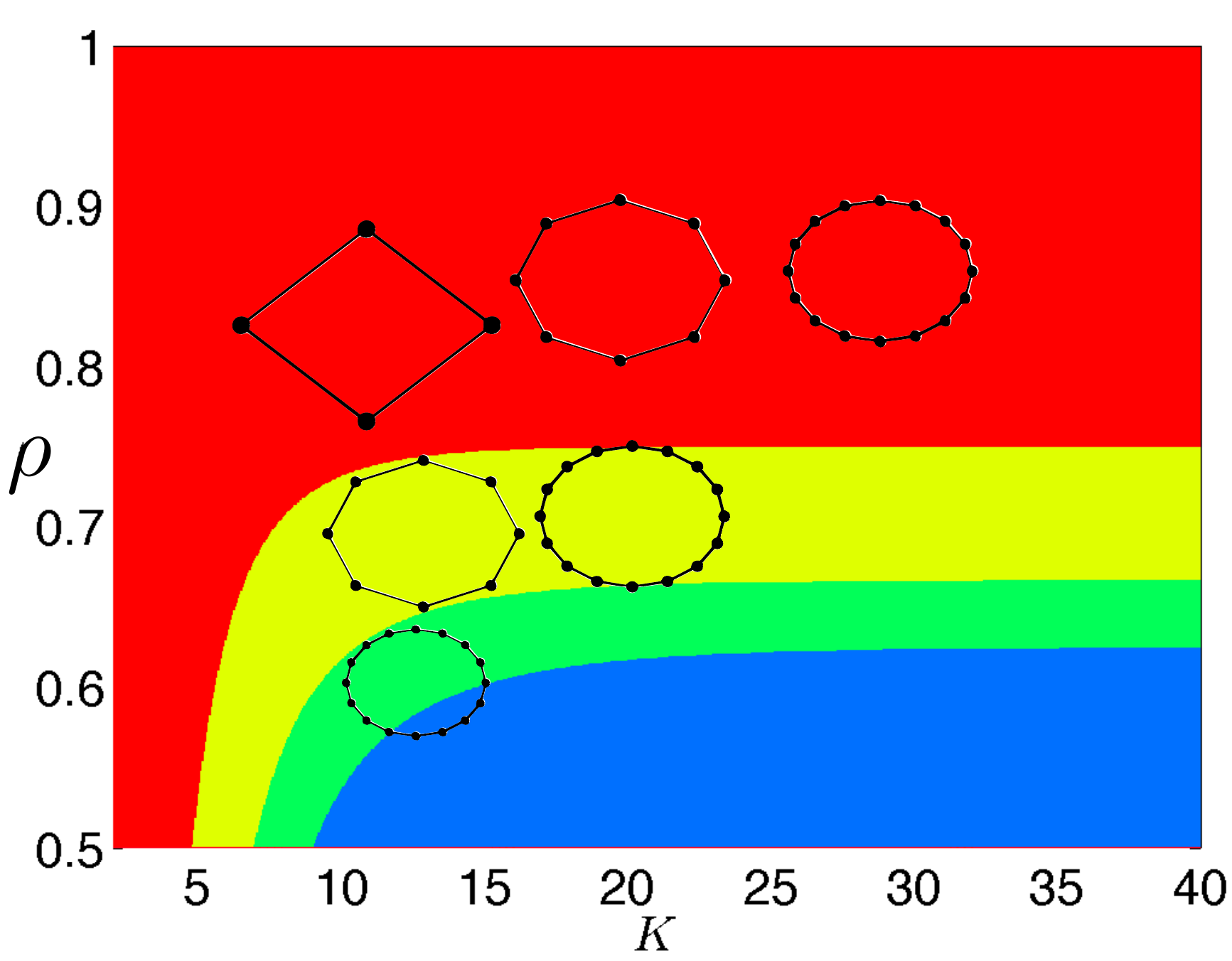}
\caption{Stability and instability zones for various configurations in the plane ($\rho$,$k$) when $\beta\rightarrow 0$, obtained by solving the inequality $S_i h_i(\sigma,k,[\textbf{S}])>0$. In particular in the figure, the square represents the configuration $S_i=+1$ $\forall i\in[1,4]$ and $S_i=-1$ $\forall i\in[5,2^{k+1}]$, the octagon the configuration $S_i=+1$ $\forall i\in[1,8]$ and $S_i=-1$ $\forall  i\in[9,2^{k+1}]$, and the esadecagon the configurationt $S_i=+1$ $\forall i\in[1,16]$ and $S_i=-1$ $\forall  i\in[17,2^{k+1}]$. In red we can see the region where all of them are stable, in yellow the region where only the octagon and the esadecagon are stable, in green the region where only the esadecagon is stable, while in blue none of these reticular animals is stable.}
\end{figure}

\subsection{Signal analysis for the Hopfield Hierarchical model}
Let us now consider the Hopfield hierarchical model (see eq.$29$). As we are interested in obtaining an explicit prescription for the fields experienced by the neurons, we can rewrite its Hamiltonian in terms of neural distance $d_{ij}$ as
\be\label{eq:2}
H_{k+1}(S|\xi,\sigma)=\sum_{i<j}S_i S_j\Big[\sum_{l=d_{ij}}^{k+1}(\frac{-1}{2^{2\sigma l}}) \Big]\sum_{\mu=1}^p \xi_i^\mu \xi_j^\mu
\ee
or inverting the order of the sums
$$
H_{k+1}(S|\xi,\sigma)=-\sum_{\mu=1}^{p} \sum_{i=1}^{2^{k+1}} S_i  \Big[\sum_{l=\mu}^{k+1}(\frac{1}{2^{2\sigma}})^l\Big] \sum_{\{j\}:d_{ij}=\mu} S_j \sum_{\nu=1}^p \xi_i^\nu \xi_j^\nu,
$$
such that, paying attention to the fields we can write
\begin{eqnarray}\label{eq:3}
H_{k+1}(S|\xi,\sigma)&=&- \sum_{i=1}^{2^{k+1}} S_i h_i[\mathbf{S}],  \\
h_i[\mathbf{S}]&=& \sum_{\mu=1}^{p} \Big[\sum_{l=\mu}^{k+1}(\frac{1}{2^{2\sigma}})^l\Big]\sum_{\{j\}:d_{ij}=\mu} S_j \sum_{\nu=1}^p \xi_i^\nu \xi_j^\nu.
\end{eqnarray}
Mirroring the analysis carried on for the Dyson model, we introduce an ensemble of non-independent Mattis-like order parameters as
\be \label{eq:4b}
m_i^{\mu,n}[\mathbf{S}]=\frac{1}{2^n} \sum_{j=i\times 2^n-(2^n-1)}^{i\times 2^n}S_j \xi_j^\mu\ \ \ \mbox{with } i=1,2,...,2^{k+1-n}, \ \ \ \mbox{ } n=0,1,2,...,k+1
\ee
so that
$$
\begin{cases}
m_i^{\mu,0}=S_i\xi_i^\mu & \mbox{with } i=1,2,..,2^{k+1}  \\
m_i^{\mu,1}=\frac{1}{2}\sum_{j=2i-1}^{2i}S_j\xi_j^\mu & \mbox{with } i=1,2,..,2^{k} \ \to m_1^{\mu,1}=\frac{1}{2}\sum_{j=1}^{2}S_j\xi_j^\mu \\
m_i^{\mu2n}=\frac{1}{2^2}\sum_{j=2^2i-(2^2-1)}^{2^2i}S_j\xi_j^\mu & \mbox{with } i=1,2,..,2^{k-1} \ \to m_1^{\mu,2}=\frac{1}{4}\sum_{j=1}^{4}S_j\xi_j^\mu \\
..... \\
m_1^{\mu,k+1}=\frac{1}{2^{k+1}}\sum_{j=1}^{2^{k+1}}S_j\xi_j^\mu.
\end{cases}
$$

As we saw for the Dyson case, this allows writing the fields as
\be\nonumber
h_i[\mathbf{S}]=\sum_{\nu=1}^p \xi_i^\nu \sum_{d=1}^{k+1} \Big[\sum_{l=d}^{k+1}(\frac{1}{2^{2\sigma}})^l\Big]2^{d-1} m_{f(d,i)}^{\nu,d-1}=\sum_{\nu=1}^p \xi_i^\nu \sum_{d=1}^{k+1} J(d,k+1,\sigma)2^{d-1}m_{f(d,i)}^{\nu,d-1},
\ee
where
\be
J(d,k+1,\sigma)2^{\mu-1}=\frac{4^{\sigma -d  \sigma }-4^{-k \sigma -\sigma}}{4^{\sigma }-1}2^{d-1}.
\ee
The microscopic evolution of the system is defined as a stochastic alignment to local field $h_{i}[\mathbf{S}]$:
\be
S_i(t+\delta t)=\textrm{sign}\{\textrm{tanh}[\beta h_i[\mathbf{S(t)}]]+\eta_i(t)\},
\ee
where the stochasticity lies in the independent random numbers $\eta_i(t)$ uniformly drawn over the interval $[-1,1]$.
In the noiseless limit $\beta \rightarrow \infty$ we have
\be
S_i(t+\delta t)=\textrm{sign}[h_i[\mathbf{S(t)}]]
\ee
and so if $S_i h_i[\mathbf{S}] >0$ $\forall \ i \ \in \ [1,N]$, the configuration $[ \mathbf{S}]$ is dynamically stable.

\subsection{Signal to noise analysis for serial retrieval}

Using equations (\ref{eq:3}) and (\ref{eq:4b}) and posing $S_i=\xi_i^\mu $ in order to check the robustness of the serial pure-state retrieval (of the test pattern $\mu$), we can write
\begin{eqnarray}
\xi_i^\mu h_i[\mathbf{S}]&=&\xi_i^\mu \sum_{\nu=1}^p \xi_i^\nu \sum_{d=1}^{k+1} J(d,k+1,\sigma)\sum_{j: d_{ij}=d}\xi_j^\nu \xi_j^\mu,\\
&=&\sum_{d=1}^{k+1} J(d,k+1,\sigma)2^{d-1}+\xi_i^\mu \sum_{\nu\neq\mu}^p \xi_i^\nu \sum_{d=1}^{k+1} J(d,k+1,\sigma)\sum_{j: d_{ij}=d}\xi_j^\nu \xi_j^\mu\nonumber.
\end{eqnarray}
We can decompose the previous equation into two contributions, a stochastic noisy term $R(\xi)$ and a deterministic signal $I$ as
\be
\xi_i^\mu h_i[\mathbf{S}]=I+R(\xi)
\ee
The signal term $I$ is positive because
\be
I=\sum_{d=1}^{k+1} J(d,k+1,\sigma)2^{d-1}\geq 0,
\ee
while the noise $R(\xi)$ has null average (the latter being denoted by standard brackets), namely
\begin{eqnarray}
R(\xi)&=&\xi_i^\mu \sum_{\nu\neq\mu}^p \xi_i^\nu \sum_{d=1}^{k+1} J(d,k+1,\sigma)\sum_{j: d_{ij}=d}\xi_j^\nu \xi_j^\mu,\\
\langle R(\xi) \rangle_\xi &=& 0.
\end{eqnarray}
Thus, in order to see the regions of the tunable parameters $\sigma,k+1$ where the signal prevails over the noise and the network accomplishes retrieval, we need to calculate the second moment of the noise over the distribution of quenched variables $\xi$ so to compare the signal amplitudes of $I$ and $|\sqrt{\langle R^2(\xi) \rangle_{\xi}}|$:

\begin{eqnarray}
\langle R^2(\xi) \rangle_\xi= \Big\langle\Big[  \sum_{\nu\neq\mu}^p \xi_i^\nu \sum_{d=1}^{k+1} J(d,k+1,\sigma)\sum_{j: d_{ij}=d}\xi_j^\nu \xi_j^\mu  \Big]\times\nonumber\\\times\Big[ \sum_{\eta\neq\mu}^p \xi_i^\eta \sum_{d=1}^{k+1} J(d,k+1,\sigma)\sum_{j: d_{ij}=d}\xi_j^\eta \xi_j^\mu  \Big]\Big\rangle_\xi.
\end{eqnarray}

Neglecting off-diagonal terms (as they have null average), we get the following expressions for $\langle R^2(\xi) \rangle_{\xi}$:
\begin{eqnarray}\small
\langle R^2(\xi) \rangle_\xi&=&\Big\langle  \sum_{\nu\neq\mu}^p (\xi_i^\nu)^2 \bigg(\sum_{d=1}^{k+1} J(d,k+1,\sigma)\sum_{j: d_{ij}=d}\xi_j^\nu \xi_j^\mu \bigg)^2 \Big\rangle_\xi=\\
&=&\left \langle  \sum_{\nu\neq\mu}^p  \bigg(\sum_{d=1}^{k+1} (\frac{4^{\sigma -d  \sigma }-4^{-({k+1}) \sigma }}{4^{\sigma }-1})\sum_{j: d_{ij}=d}\xi_j^\nu \xi_j^\mu \bigg)^2 \right \rangle_\xi,\nonumber
\end{eqnarray}
where we used $(\xi_i^{\nu})^2=1$ $\forall i,\nu$.
Once again,  as the $\xi$'s are symmetrically distributed, only even order terms give contributions, thus we can safely  neglect off-diagonal terms and write again
\begin{eqnarray}
\langle R^2(\xi) \rangle_\xi &=&(p-1)\sum_{d=1}^{k+1} \left \langle    \left[ \left(\frac{4^{\sigma -d  \sigma }-4^{-k \sigma -\sigma}}{4^{\sigma }-1} \right)\sum_{j: d_{ij}=d}\xi_j^\nu \xi_j^\mu \right]^2  \right\rangle_\xi, \\
 &=&(p-1)\sum_{d=1}^{k+1}    \bigg(\frac{4^{\sigma -d  \sigma }-4^{-k \sigma-\sigma }}{4^{\sigma }-1}\bigg)^2  \langle \sum_{j: d_{ij}=d,}\sum_{k: d_{ik}=d}\xi_j^\nu \xi_j^\mu \xi_k^\nu \xi_k^\mu \rangle_\xi.\nonumber
\end{eqnarray}
Therefore
\be
\langle R^2(\xi) \rangle_\xi=(p-1)\sum_{d=1}^{k+1}    J(d,\sigma,k+1)^2 2^{d-1}.
\ee
Exploiting the approximation  $\langle |x| \rangle \sim |\sqrt{\langle x^2 \rangle}|$, we can simplify the previous expression into
\be
\langle |R(\xi)|\rangle \sim \sqrt{\langle R^2(\xi) \rangle_\xi}=\sqrt{(p-1)\sum_{d=1}^{k+1}   J(d,\sigma,k+1)^2 2^{d-1}},
\ee
where we consider the positive branch of the serial retrievl only. We are now ready to check the stability of the pure retrieval: as long as
\be
 I>\sqrt{\langle R^2(\xi) \rangle_\xi} \Rightarrow\xi_i^\mu h_i[\mathbf{S}]=I+R(\xi)>0,
\ee
the pure state is stable. Hence we need to calculate explicitly
\be
\nonumber
\sqrt{ \langle R^2(\xi) \rangle_\xi} =
\sqrt{\frac{(p-1) 16^{-k \sigma } }
{\left(4^{\sigma }-2\right) \left(4^{\sigma }-1\right)^2 \left(16^{\sigma }-2\right)}}\cdot \sqrt{\Psi_1 + \Psi_2},
\ee
where
\begin{eqnarray}\nonumber
 \Psi_1 &=& (4^{\sigma }-2 ) 4^{2 (k+1) \sigma }-3\times 2^{k+2 \sigma +1},\\ \nonumber
 \Psi_2 &=& 2^{k+6 \sigma +1}-(16^{\sigma }-2 )
2^{2 (k+1) \sigma +1}+2^{k+2}-64^{\sigma }+2^{2 \sigma +1}+2^{4 \sigma +1}-4.
\end{eqnarray}
The expression for the signal is much simpler, resulting in
\be
 I=\frac{4^{-(k+1) \sigma } \left(-2^{k+2 \sigma +2}+4^{(k+2) \sigma }+2^{k+2}+4^{\sigma }-2\right)}{-3 \times 4^{\sigma }+16^{\sigma }+2}.
\ee
Imposing $I=\sqrt{\langle R^2(\xi) \rangle_\xi}$ and solving for the variable $p$, we find the critical load allowed by the network, namely the function $P_{c}(\sigma,k)$, whose behavior is shown in Fig.$5$:
\be
I=\sqrt{\langle R^2(\xi) \rangle_\xi}\Rightarrow P_{c}(\sigma,k).
\ee

Now, imposing the relation
$$
P_{c}(\sigma,k)=k
$$
and solving numerically with respect to $\sigma$, we can plot the maximum value $\sigma_{\max}(k)$ that the variable $\sigma$ can reach such that the storage $P=k$ produces retrievable patterns, as shown in Figure $5$.
\newline
In the thermodynamic limit we get
\begin{eqnarray}
I-\sqrt{\langle R^2(\xi)\rangle}&=&\frac{2^{2\sigma}}{-3 \times 4^\sigma+16^\sigma+2}-\frac{\sqrt{(p-1)} 2^{2\sigma}}{\sqrt{\left(4^\sigma-1\right) \left(16^\sigma-2\right)}},\\
P_{c}(\sigma)&=&\frac{\left(4^\sigma-1\right) \left(16^\sigma-2\right)}{\left(-3\times  4^\sigma+16^\sigma+2\right)^2}+1.
\end{eqnarray}

\subsection{Signal to noise analysis for parallel retrieval }

Fixing  $S_i=\xi_i^\mu $ $\forall i \in[1,2^{k}]$ and $S_i=\xi_i^\gamma $ $\forall i \in[1+2^{k},2^{k+1}]$ for $\mu\neq\gamma$, namely selecting $\mu$ and $\gamma$ as test patterns to retrieve, we set the system in condition to handle contemporarily two patterns, the former managed by the first half of the neurons, the latter by the second half. The robustness of this state is addressed hereafter following the same prescription outlined so far. Namely, being
\be
S_i h_i[\mathbf{S}]=S_i \sum_{\nu=1}^p \xi_i^\nu \sum_{d=1}^{k+1} J(d,k+1,\sigma)\sum_{j: d_{ij}=d}\xi_j^\nu S_j,
\ee
if $i\in[1,2^{k}]$ we have
\be
\nonumber
 S_i h_i (S) = \xi_i^\mu \sum_{\nu=1}^p \xi_i^\nu \bigg( \sum_{d=1}^{k} J(d,k+1,\sigma)\sum_{j: d_{ij}=d}\xi_j^\nu \xi_j^\mu + J(k+1,k+1,\sigma)\sum_{j: d_{ij}=k+1}\xi_j^\nu \xi_j^\gamma \bigg),
\ee
while if $i \in[2^{k}+1,2^{k+1}]$, the same equation still holds provided we replace $\mu$ with $\gamma$ and $\gamma$ with $\mu$, hence hereafter we shall consider only one of the two cases as they are symmetrical.
Again, we can decompose the above expression in the sum of a constant, positive term -that plays as the signal- $I>0$, and a stochastic term for the noise $R(\xi)$, namely we can write
\begin{eqnarray}
S_i h_i[S]&=&I+R(\xi),\\ \nonumber \small
I&=&\sum_{d=1}^{k}\bigg( J(d,k+1,\sigma)2^{d-1} \bigg),\\ \nonumber \small
R(\xi)&=&J(k+1,k+1,\sigma)\sum_{j: d_{ij}={k+1}}\xi_j^\mu \xi_j^\gamma \\ \nonumber \small
 &+& \xi_i^\mu \sum_{\nu\neq\mu}^p \xi_i^\nu \big( \sum_{d=1}^{k} J(d,k+1,\sigma)\sum_{j: d_{ij}=d}\xi_j^\nu \xi_j^\mu + J(k+1,k+1,\sigma)\sum_{j: d_{ij}=k+1}\xi_j^\nu \xi_j^\gamma \big).
\end{eqnarray}

In order to get a manageable expression for the noise, it is convenient to reshuffle $R(\xi)$ distinguishing four terms such that
\be
R(\xi)=a+b+c+d,
\ee
where
\begin{eqnarray}
a&=& J(k+1,k+1,\sigma)\sum_{j: d_{ij}=k+1}\xi_j^\mu \xi_j^\gamma,\\
b&=& \xi_i^\mu \sum_{\nu\neq\mu}^p \xi_i^\nu  \sum_{d=1}^{k} J(d,k+1,\sigma)\sum_{j: d_{ij}=d}\xi_j^\nu \xi_j^\mu,\\
c&=& \xi_i^\mu \sum_{\substack{\nu\neq\mu \\ \nu\neq\gamma}}^p \xi_i^\nu  J(k+1,k+1,\sigma)\sum_{j: d_{ij}=k+1}\xi_j^\nu \xi_j^\gamma,\\
d&=& \xi_i^\mu \xi_i^\gamma J(k+1,k+1,\sigma)2^{k}.
\end{eqnarray}
As $\mu\neq\gamma$, we have that $\langle R(\xi) \rangle_\xi=0$, while $\langle R^2(\xi) \rangle_\xi$ turns out to be
\be
\langle R^2(\xi) \rangle_\xi=\langle a^2+b^2+c^2+d^2+2(ab+ac+ad+bc+bd+cd)\rangle_\xi.
\ee
Let us consider these terms separately: skipping lenghty, yet straightforward calculations, we obtain the following expressions
\begin{eqnarray}\nonumber
\langle a^2 \rangle_\xi&=&\Big\langle J^2(k+1,k+1,\sigma)\sum_{j: d_{ij}=k+1}\sum_{n: d_{in}=k+1}\xi_j^\mu \xi_j^\gamma \xi_n^\mu \xi_n^\gamma  \Big\rangle_\xi\\
&=& J^2(k+1,k+1,\sigma) \times 2^{k}.
\end{eqnarray}
\begin{eqnarray}\nonumber
\langle b^2 \rangle_\xi &=& \left\langle \bigg(\xi_i^\mu \sum_{\nu\neq\mu}^p \xi_i^\nu  \sum_{d=1}^{k} J(d,k+1,\sigma)\sum_{j: d_{ij}=d}\xi_j^\nu \xi_j^\mu \bigg)^2 \right\rangle_\xi\\
&=& (p-1) \sum_{d=1}^{k} J^2(d,k+1,\sigma) 2^{d-1}.
\end{eqnarray}
\begin{eqnarray}\nonumber
\langle c^2 \rangle_\xi &=& \left \langle \bigg( \xi_i^\mu \sum_{\nu\neq\mu \& \nu\neq\gamma}^p \xi_i^\nu  J(k+1,k+1,\sigma)\sum_{j: d_{ij}=k+1}\xi_j^\nu \xi_j^\gamma  \bigg)^2 \right\rangle_\xi\\
&=& (p-2)  J^2(k+1,k+1,\sigma)2^{k}.
\end{eqnarray}
\be
\langle d^2 \rangle_\xi = \left\langle \bigg( \xi_i^\mu \xi_i^\gamma J(k+1,k+1,\sigma)2^{k} \bigg)^2 \right\rangle_\xi
=J^2(k+1,k+1,\sigma)2^{2k},
\ee

and, since $a$ and $b$ and, analogously, $b$ and $c$, are defined over different blocks of spins, clearly
\begin{eqnarray}
\langle 2ab \rangle_\xi &=& 0,\\
\langle 2bc \rangle_\xi &=& 0,\\
\langle 2bd \rangle_\xi &=& 0.
\end{eqnarray}
As a result, rearranging terms opportunely we finally obtain
\begin{eqnarray}\nonumber
\langle R^2(\xi) \rangle_\xi &=& 4^{-2 k \sigma}\Big(\frac{ \left[4^k \left(4^{\sigma }-1\right)^2+2^{k} \left(4^{\sigma }-1\right)^2+2^{k} (p-2) \left(4^{\sigma }-1\right)^2\right ]}{\left(4^{\sigma }-1\right )^2}\\
\nonumber
&+&   (2((-3\times 2^{k+2 \sigma+1}+2^{k+6 \sigma+1}+2^{k+2}+2^{2 \sigma+1}+2^{4 \sigma+1}-\\ \nonumber
&+& (4^{\sigma}-2) 4^{2 (k+1) {\sigma}}-(16^{{\sigma}}-2) 2^{2 (k+1) {\sigma}+1})+\\
   \nonumber
   &-& 64^{\sigma})(p-1))((4^{\sigma}-2)(16^{\sigma}-2))^{-1}\Big),
\end{eqnarray}

while the signal term reads as
\be
I=\frac{2^{-2 k \sigma -1} \left(-2^{k+2 \sigma }-2^{k+4 \sigma }+2^{2 (k+1) \sigma +1}+2^{k+1}+2^{2 \sigma +1}-4\right)}{-3\times 4^{\sigma }+16^{\sigma }+2}.
\ee
Imposing $I=\sqrt{\langle R^2(\xi) \rangle_\xi}$, and solving with respect to the variable $p$ we can outline the function $P_{c}(\sigma,k+1)$ that returns the maximum allowed load the network may afford accomplishing parallel retrieval and whose behavior is shown in Fig.$5$:
\be
I=\sqrt{\langle R^2(\xi) \rangle_\xi}\Rightarrow P_{c}(\sigma,k+1).
\ee


\begin{figure}[hd] \label{fig6}
\centering
\includegraphics[width=6.5cm]{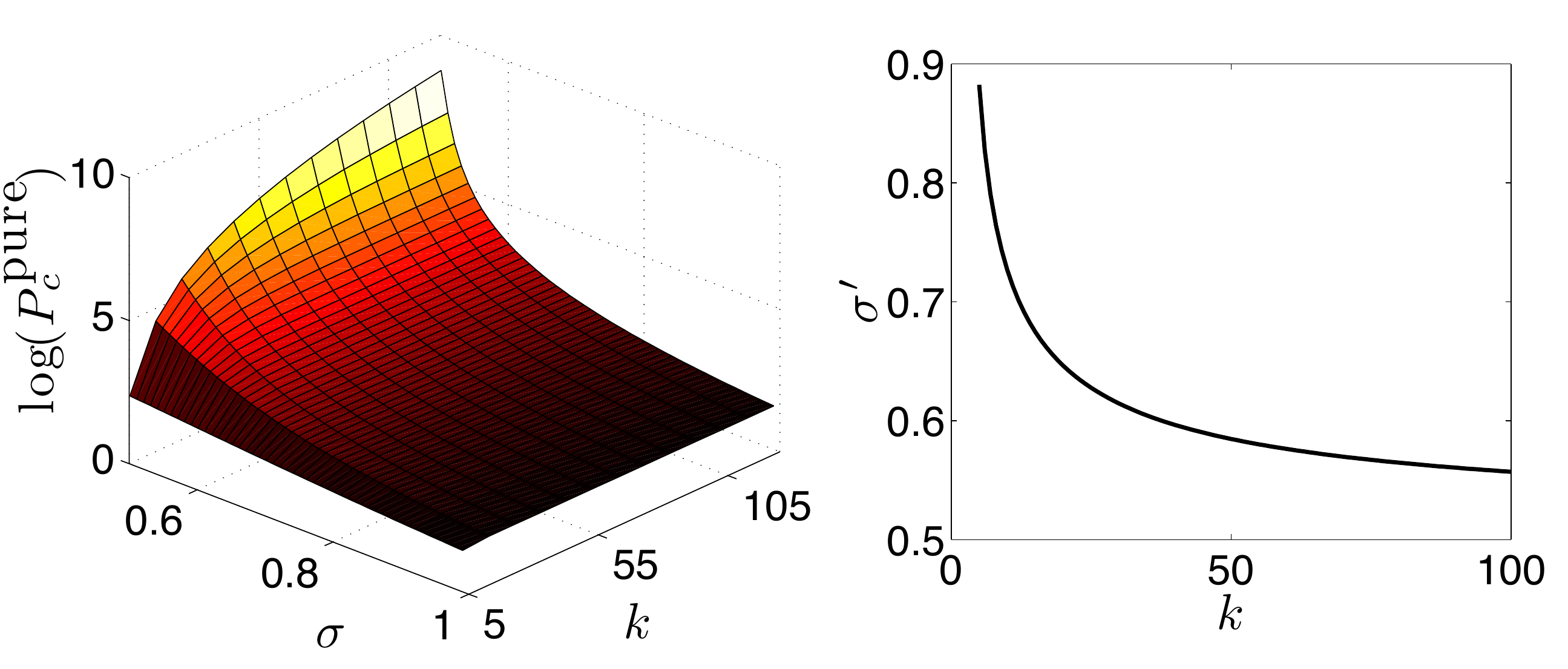}\\
\includegraphics[width=6.5cm]{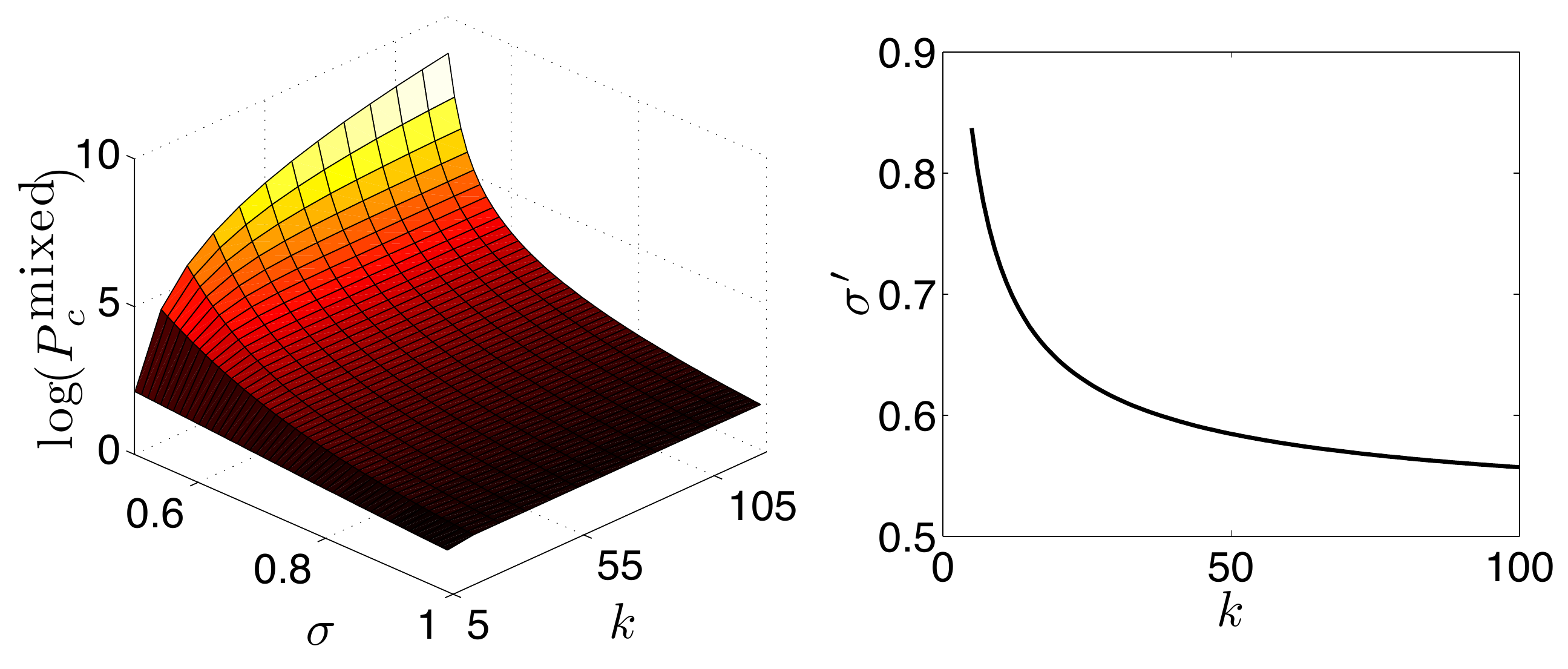}
\caption{Upper panel (serial retrieval): On the left we show the maximum value of storable patterns $P_c$ as a function of $k$ and of $\sigma$ (as results from Eq.~$72$) for the pure state in order to have signal's amplitude greater than the noise (i.e. retrieval). Note the logarithmic scale for $P_c$ highlighting its wide range of variability. On the right we show the maximum value of the neural interaction decay rate $\sigma'(k)$ versus $k$ allowed to the couplings under the storage constraint $k = p$ and the pure state perfect retrieval constraint, in the $\beta \to \infty$ limit.\newline
Lower panel (parallel retrieval): On the left there is the maximum value of storable patterns $P_c$ as a function of $k$ and of $\sigma$ (as results from Eq.~$91$) for th parallel state in order to have signal's amplitude greater than the noise (i.e. retrieval). Note the logarithmic scale for $P_c$ highlighting its wide range of variability. On the right there is the maximum value of the neural interaction decay rate $\sigma'(k)$ versus $k$ allowed to the couplings under the storage constraint $k = p$ and the parallel state perfect retrieval constraint, in the $\beta \to \infty$ limit.}
\end{figure}

\begin{figure}[tb] \begin{center}
\includegraphics[width=1.00\textwidth]{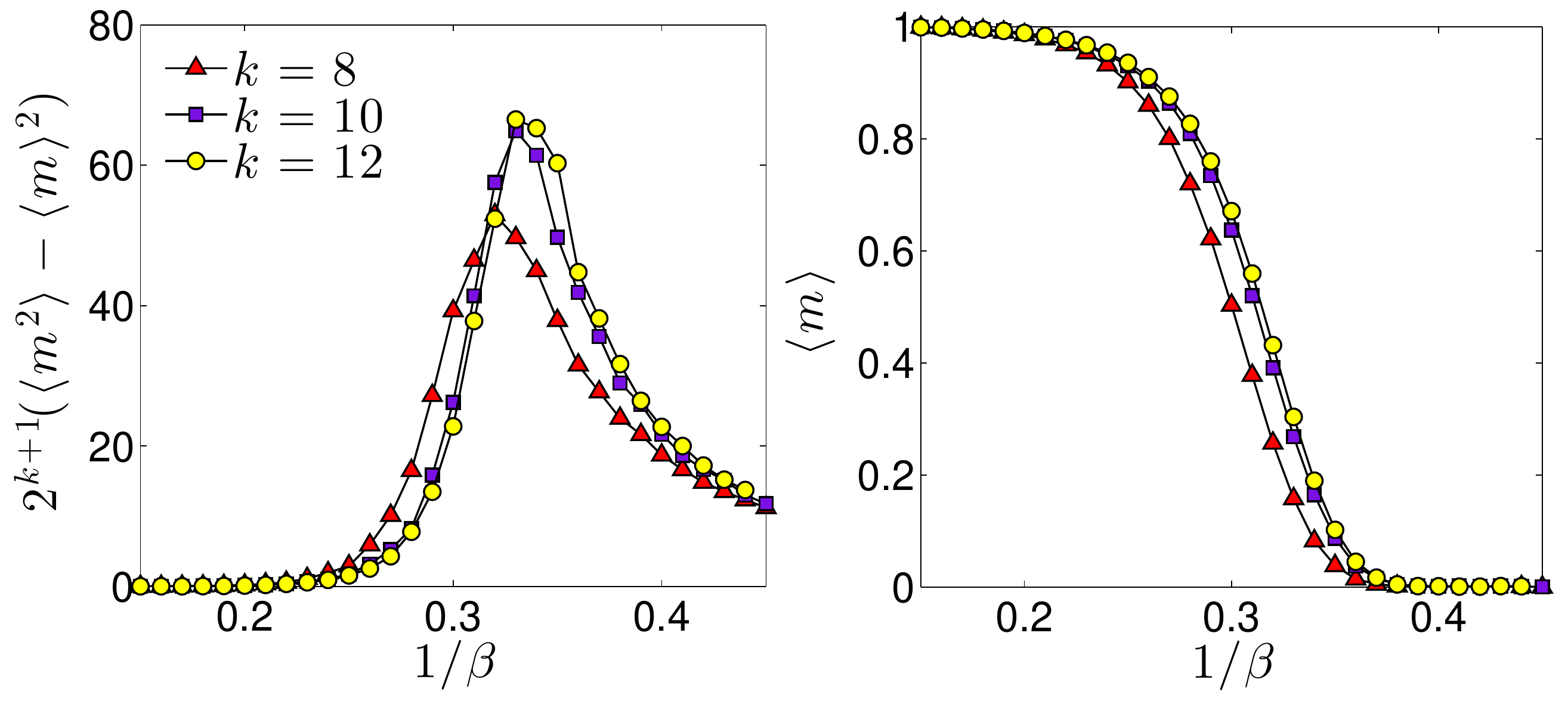}
\caption{\label{fig:seriale}
Starting from the state $S_i=+1$ $\forall i \in[1,2^{k+1}]$ results of the simulations for DHM for $\sigma=0.99 $ and $N=2^{k+1}$, $k+1=8,10,12$ are plotted. In the left panel, the rescaled magnetic susceptibility $2^{k+1}(\langle m^2\rangle -\langle m\rangle^2) $ is plotted vs $\beta$ (one over the noise).
In the right panel the magnetization $\langle m\rangle =\langle \frac{1}{N}\sum_{i=1}^{N}S_i\rangle$ is plotted vs $\beta$ (one over the noise).}
\end{center}
\end{figure}

\begin{figure}[tb] \begin{center}
\includegraphics[width=1.00\textwidth]{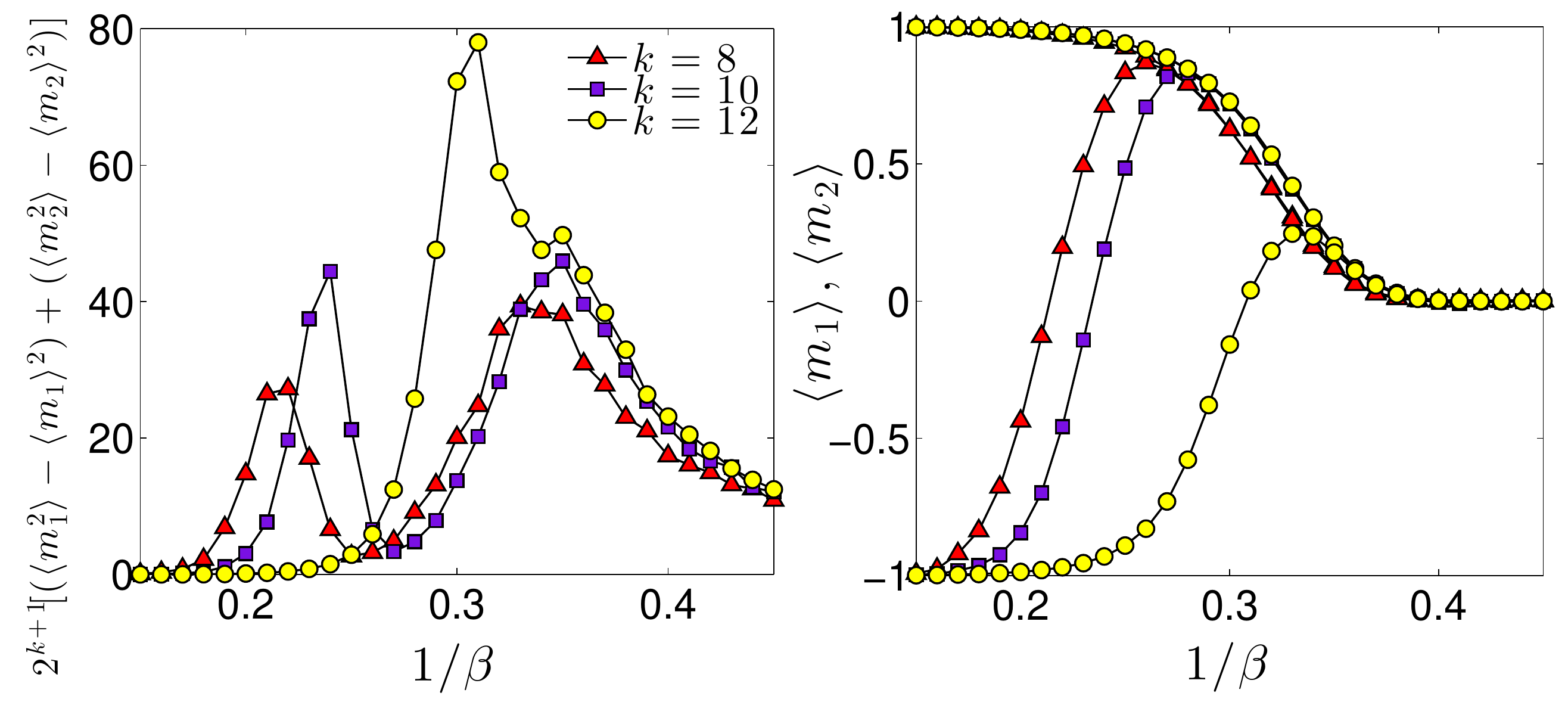}
\caption{\label{fig:parallelo}
Starting from the state $S_i=+1,S_j=-1$ $\forall i \in[1,2^{k}]$ and $\forall j \in[2^{k}+1,2^{k+1}]$ results of the simulations for DHM for $\sigma=0.99 $ and $N=2^{k+1}$ are plotted. In the left panel, the rescaled magnetic susceptibility $2^{k+1}[(\langle m_1^2\rangle -\langle m_1\rangle ^2)+(\langle m_1^2\rangle-\langle m_1\rangle^2] $ is plotted vs $\beta$ (i.e. one over the noise) for $k+1=8,10,12$.
In the right panel, the magnetizations $\langle m_1\rangle =\langle \frac{1}{2^k}\sum_{i=1}^{2^k}S_i\rangle $ and $\langle m_2\rangle=\langle\frac{1}{2^k}\sum_{i=1+2^k}^{2^{k+1}}S_i\rangle$  are plotted vs $\beta$ (i.e. one over the noise) for $k+1=8,10,12$.}
\end{center}
\end{figure}

\newpage
\subsection{Insights from numerical simulations} \label{sec:N}

Aim of this Section is to present results from extensive numerical simulations to check the stability of parallel processing over the finite-size effects that is not captured by statistical mechanics or that can be hidden in the signal-to-noise analysis . Further this allows checking that the asymptotic behavior (in the volume) of the network is in agreement with previous findings.
\newline
All the simulations were carried out according to the following algorithm.
\begin{itemize}
\item[1.]  Building the matrix coupling, pattern storage.\\
Once extracted randomly from a uniform prior over $\pm 1$ $p$ patterns of length $k+1$, and defined the distance between two spins  $i$ and $j$ as $d_{ij}$ we build the matrix $\mathbf{J}$, for the HHM, as
\be
J_{ij}=\frac{4^{\sigma-d_{ij}\sigma}-4^{-(k+1)\sigma}}{4^{\sigma}-1}\sum_{\mu=1}^p \xi_i^{\mu}\xi_j^{\mu}, \text{ for } i=1,\cdots 2^{k+1},\text{  }j=1,\cdots,2^{k+1},
\ee
while for the DHM we use the form:
\be
J_{ij}=\frac{4^{\sigma-d_{ij}\sigma}-4^{-(k+1)\sigma}}{4^{\sigma}-1}, \text{ for } i=1,\cdots 2^{k+1}\text{ and }j=1,\cdots,2^{k+1},
\ee
where $k+1$ is the number of levels of the hierarchical construction of the network, and $\sigma\in(\frac{1}{2},1]$.

\item[2.] Initialize the network.\\
We used different initializations to test the stability of the resulting stationary configuration:

-Pure retrieval: We initialize the network in an assumed fixed point of the dynamics, namely $S_i=\xi_i^{\mu}$ with $i=1,...2^{k+1}$ and $\mu=1$ for the HHM, while $S_i=+1$ with $i=1,...2^{k+1}$ in the DHM case, and we check the equilibrium as reported in Fig[~\ref{fig:seriale}].

-Parallel retrieval: Since we study the multitasking features shown by this hierarchical network, we can also assign different types of initial conditions with respect to the pure state, e.g.

\begin{itemize}
\item[i)] For the DHM, starting from the lowest energy level ( after the standard one $S_i=1$ $\forall i$) we chose $S_i=+1$ for $i=1,...,2^{k}$ and $S_i=-1$ for $i=2^{k}+1,...,2^{k+1}$ (viceversa is the same, and we check the equilibrium  as reported in Fig[~\ref{fig:parallelo}]);
\item[ii)] For the HHM, looking for multitasking features, we set in the case $p=2$, we set $S_i=\xi_i^1$ for $ i=1,...,2^{k}$ and $S_i=\xi_i^2$  $i=2^{k}+1,...,2^{k+1} $(Fig[~\ref{fig:p2}]); In the case $p=4$ we set $S_i=\xi_i^\mu$ $\forall i \in \big[1+\frac{(\mu-1)N}{4},\frac{\mu N}{4}\big]$ and $\mu \in [1,4] $(Fig[~\ref{fig:p4}])
\end{itemize}

In this way, we have two or four communities (sharing the same size) building the network with a different order parameter.

\item[3.] Evolution: Glauber dynamics.\\
 The evolution of the spins follows a standard random asynchronous dynamics \cite{peter} and the state of the network is updated according to the field acting on the spins at every step of iteration, that is,
\be
\nonumber
S_i(t+1)=\sign\{\tanh[\beta h_i(\mathbf{S}(t)]+\eta(t)\},\text{ for }\, \beta=T^{-1}
\ee
where $\eta(t)$ is the noise introduced as a random uniform contribution over the real interval $[-1,1]$ in every step.
\newline
For each noise the stationary mean values of the order parameters have been measured mediating over $O(10^3)$ different realizations.
For the HHM the average of the order parameters is  performed over the quenched variables.
For DHM, to better highlight the stability of the parallel configuration, $S_i=+1$ for $i=1,...,2^{k}$, $S_i=-1$ for $i=2^{k}+1,...,2^{k+1}$ and to break the Gauge invariance, during half of the relaxation period  to equilibrium a small positive field is applied to the system.

\begin{figure}[h!] \begin{center}
\includegraphics[width=.75\textwidth]{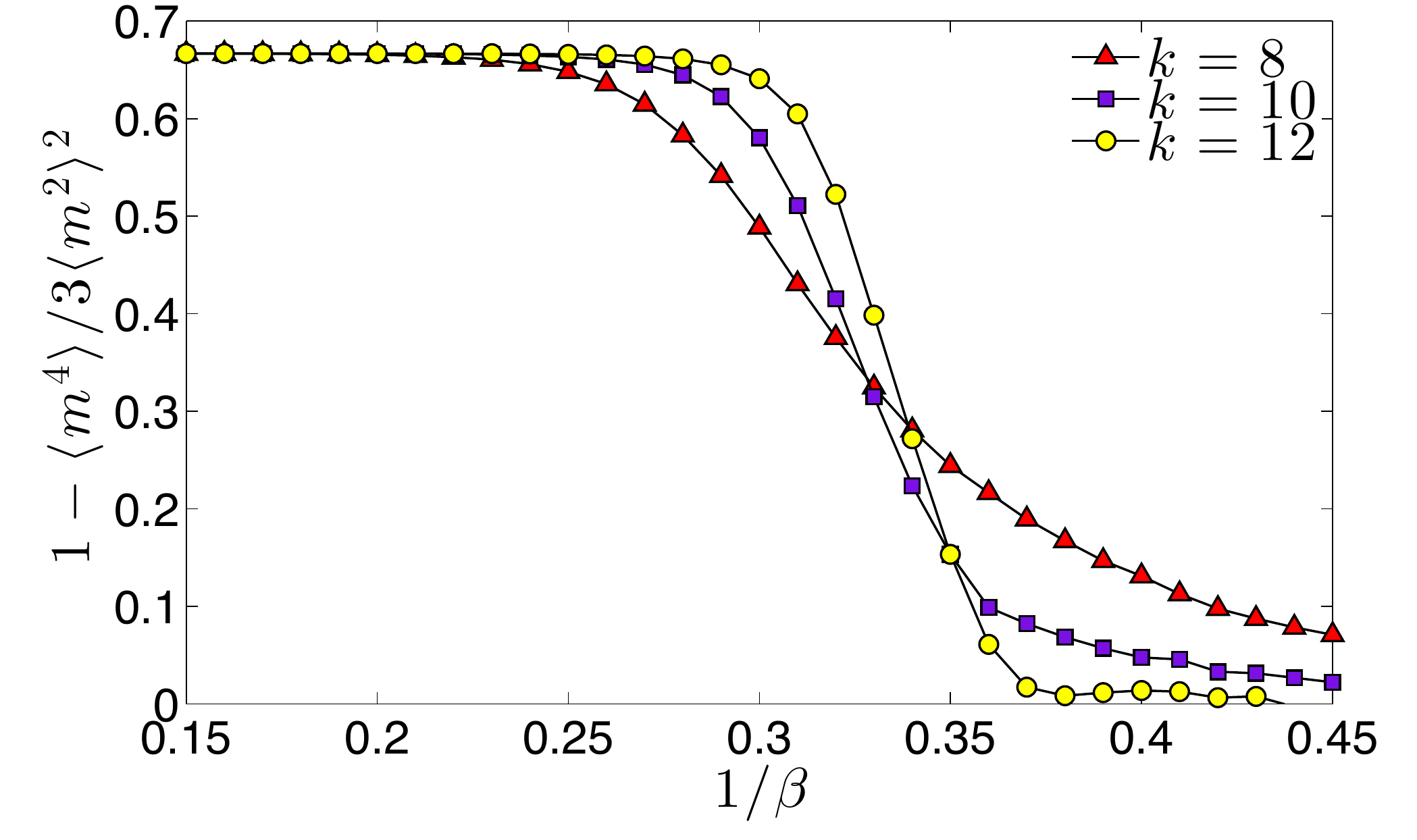}
\caption{\label{fig:binder} Starting from the state $S_i=+1\text{ } \forall i \in[1,2^{k+1}]$ with $\sigma=0.99 $ for the  DHM  and $k+1=8,10,12$. Binder cumulant $1-\frac{\langle m^4\rangle}{3	\langle m^2\rangle^2}$ versus noise $\frac{1}{\beta}$ for $k+1=8,10,12$.  Plotting the binder cumulant for different values of $k+1$ permits to find the critical noise of this state.}
\end{center}
\end{figure}

\item[4.] Results.\\
It is worth noting that -at difference with paradigmatic prototypes for phase transitions (i.e. the celebrated Curie-Weiss model), as we can see from figures [~\ref{fig:seriale},~\ref{fig:parallelo},~\ref{fig:binder}], in these models we studied here the critical noise level approaches its asymptotic value (obtained by analytical arguments in the thermodynamic limit) from above (i.e. from higher values of $\beta$s).
This happens because the intensities of couplings are increasing functions (clearly upper limited) of the size of the system.
As can be inferred from fig[~\ref{fig:parallelo}] (where we present results regarding simulations for  the DHM  at $\sigma=0.99$,  $k+1=8,10,12$ [$S_i=+1,S_j=-1$ $\forall i \in[1,2^{k}]$ and $\forall j \in[2^{k}+1,2^{k+1}]$]), the stability of the parallel configuration (in the low noise region) is confirmed and, as expected from theoretical arguments, the noise region in which this configuration is stable increases with the size of the system up to coincide with that of the pure state. Also in the HHM case (figures [~\ref{fig:p4},~\ref{fig:p2}]) the stability of parallel configurations is verified (in the low noise region) for system's configurations shared by the two and four communities.

\end{itemize}

\begin{figure}[h!tb] \begin{center}
\includegraphics[width=1.00\textwidth]{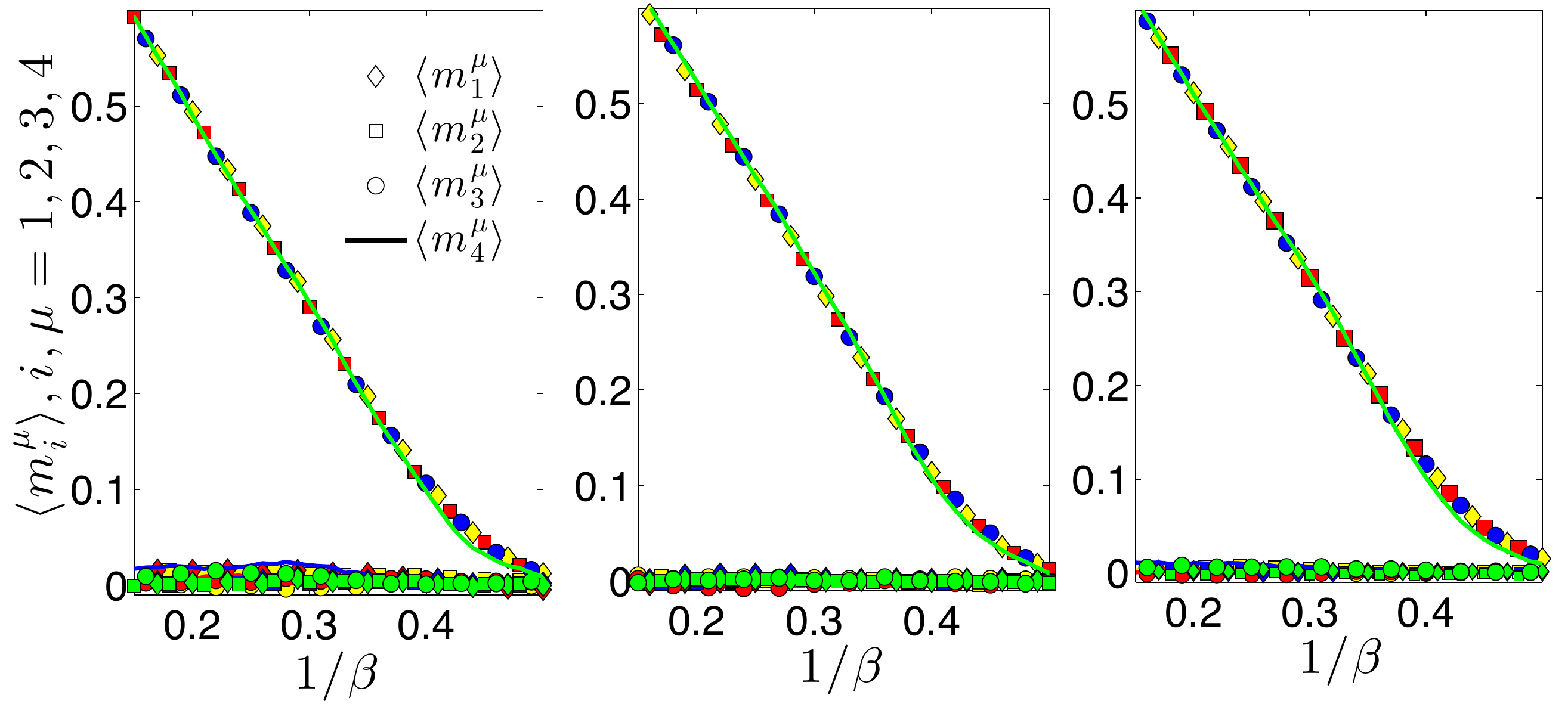}
\caption{\label{fig:p4} .
Starting from the state $S_i=\xi_i^1,S_j=\xi_j^2,S_n=\xi_n^3,S_l=\xi_l^4$ $\forall i \in[1,2^{k-1}],\forall j \in[2^{k-1}+1,2^{k}],\forall n \in[2^k+1,\frac{3}{2}2^k], \forall l \in[\frac{3}{2}2^{k}+1,2^{k+1}]$ results of the simulations for HHM for $\sigma=0.99 $ and $N=2^{k+1}$ are plotted.
The Mattis order parameters $\langle m^\mu_i \rangle =\langle \frac{1}{2^{k-2}}\sum_{j=1+{(i-1)2^{k-2}}}^{i 2^{k-2}}S_j\xi_j^\mu\rangle$ for $i,\mu\in[1,4]$ are plotted vs noise,from left we have $k+1=8,10,12$. Same colors correspond to the same pattern $\mu$, while same symbols correspond to the same index $i$.
}
\end{center}
\end{figure}
\begin{figure}[h!tb] \begin{center}
\includegraphics[width=1.00\textwidth]{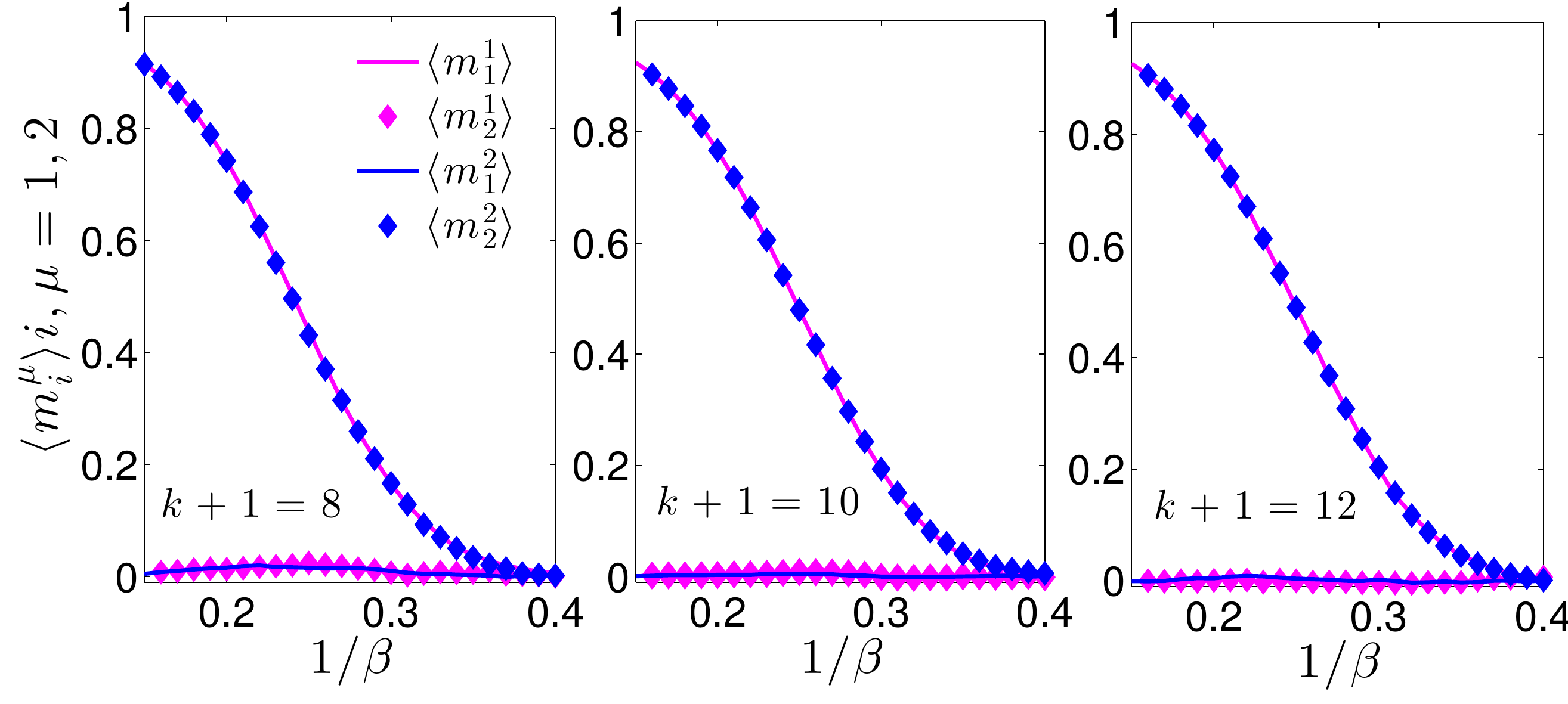}
\caption{\label{fig:p2}.
Starting from the state $S_i=\xi_i^1,S_j=\xi_j^2$ $\forall i \in[1,2^{k}],\forall j \in[2^{k}+1,2^{k+1}]$ results of the simulations for HHM for $\sigma=0.99 $ and $N=2^{k+1}$ are plotted.
The Mattis order parameters $\langle m^\mu_i \rangle =\langle \frac{1}{2^{k-2}}\sum_{j=1+{(i-1)2^{k-2}}}^{i 2^{k-2}}S_j\xi_j^\mu\rangle$ for $i,\mu\in[1,2]$ are plotted vs noise,from left we have $k+1=8,10,12$.}
\end{center}
\end{figure}

\newpage
\section{Conclusions and outlooks}

Comprehension of biological complexity is one of the main aim of this century's research: the route to pave is long and scattered over countless branches. Restricting to neural networks, due to prohibitive constraints when dealing with statistical mechanics beyond the mean field approximation (where each notion of distance or metrics for a space where to embed neurons is lost), their theory has been largely developed without investigating the crucial degree of freedom of neural distance. However, research is nowadays capable of investigations towards more realistic and/or better performing models: indeed, while the mean-field scenario, mainly split among Hopfield network for retrieval and Boltzmann machines for learning, has been so far understood (not completely at the rigorous level but at least largely), investigation of the non-mean-field counterpart is only at the beginning.
\newline
In this work we tackled the problem of studying information processing (retrieval only) on hierarchical topologies, where neurons interact with an Hebbian strength (or simply ferromagnetically in their simplest implementation, namely the Dyson model) that decays with their reciprocal distance. While a full statistical mechanical treatment is not yet achievable, stringent bounds for its free energy -intrinsically of non-mean-field nature- are however available and return a survey of network capabilities by far richer than the corresponding mean-field counterpart (the Hopfield model within the low storage regime). Indeed these network are able to retrieve one pattern at a time accomplishing an extensive reorganization of the whole neuronal state -mirroring serial processing as in standard Hopfield networks- but they are also able to switch to multitasking behavior handling multiple patterns at once -without falling into spurious states-, hence performing as parallel processors.
\newline
Remarkably, as far as the low storage regime is concerned, this defragmentation into cliques -crucial for parallel processing- returns a phase space that shares huge similarities with the multitasking associative networks \cite{PRL}.
\newline
However, as theorems that definitively confirm this scenario are not yet fully available, to give robustness to the statistical mechanics predictions, we performed a signal-to-noise analysis checking whether those states -candidate by the first approach to mimic parallel retrieval- are indeed stable beyond the pure state related to serial processing and remarkably we found huge regions of the tunable parameters (strength of the interaction decay $\sigma$ and noise level $\beta$) where indeed those states are extremely robust.
\newline
Clearly, as standard in thermodynamics, nothing is for free and even for this richness of behaviors there is a price to pay: emergent multitasking features in not-mean-field models require a substantial  drop in network's capacity  thus implying a new balance required by associative networks beyond the mean-field scenario.
\newline
While a satisfactory picture beyond such a mean-field paradigm is still far, but we hope that this work may act as one of the first steps toward this direction.

\section*{Acknowledgments}
The authors acknowledge partial financial support from the GNFM (INdAM) -Gruppo Nazionale per la Fisica Matematica- [thanks to Progetto Giovani Barra 2013 and Progetto Giovani Agliari 2014], INFN -Istituto Nazionale di Fisica Nucleare- and Sapienza Universita' di Roma.


\end{document}